\documentclass[a4paper,5p,times,sort&compress]{elsarticle}%5p for double column
\usepackage[utf8]{inputenc}
\usepackage[T1]{fontenc}

\graphicspath{{graphics/}}

\usepackage{amsmath}
\usepackage{amsfonts}
\usepackage{amssymb}
\usepackage{float}

\usepackage[normalem]{ulem}

\usepackage{booktabs}
\usepackage{tabularx}
\usepackage{threeparttable}
\usepackage{rotating}
\usepackage{siunitx}

\usepackage{url}
\usepackage[colorlinks=true, citecolor=blue, linkcolor=blue, filecolor=blue,urlcolor=blue]{hyperref}

\usepackage[labelfont=bf]{caption}
\usepackage[gen]{eurosym}

\usepackage{hyphenat}

\usepackage{lineno}
\modulolinenumbers[5]

\usepackage{framed} %

\usepackage{multicol} %

\usepackage{nomencl} %

\makenomenclature

\renewcommand*\nompreamble{\begin{multicols}{2}}

\renewcommand*\nompostamble{\end{multicols}}

\journal{Energy}

\begin{document}
%\linenumbers

\begin{frontmatter}

\title{Cost optimal scenarios of a future highly renewable European electricity system: Exploring the influence of weather data, cost parameters and policy constraints}

\author[fias,aarh]{D.P.~Schlachtberger \corref{cor1}}
\ead{schlachtberger@fias.uni-frankfurt.de}
\author[kit,fias]{T.~Brown}
\author[aarh,inatech]{M.~Sch\"{a}fer}
\author[fias]{S.~Schramm}
\author[aarh]{M.~Greiner}
\cortext[cor1]{Corresponding author}

\address[fias]{Frankfurt Institute for Advanced Studies, 60438 Frankfurt am Main, Germany}
\address[aarh]{Department of Engineering, Aarhus University, 8000 Aarhus C, Denmark}
\address[kit]{Institute for Automation and Applied Informatics, Karlsruhe Institute of Technology, Hermann-von-Helmholtz-Platz 1, 76344 Eggenstein-Leopoldshafen, Germany}
\address[inatech]{Department of Sustainable Systems Engineering (INATECH), University of Freiburg, 79110 Freiburg, Germany}

\begin{abstract}
Cost optimal scenarios derived from models of a highly renewable electricity system depend on the specific input data, cost assumptions and system constraints. Here this influence is studied using a techno\hyp{}economic optimisation model for a networked system of 30 European countries, taking into account the capacity investment and operation of wind, solar, hydroelectricity, natural gas power generation, transmission, and different storage options. A considerable robustness of total system costs to the input weather data and to moderate changes in  the cost assumptions is observed. Flat directions in the optimisation landscape around cost-optimal configurations often allow system planners to choose between different technology options without a significant increase in total costs, for instance by replacing onshore with offshore wind power capacity in case of public acceptance issues. Exploring a range of carbon dioxide emission limits shows that for scenarios with moderate transmission expansion, a reduction of around 57\% compared to 1990 levels is already cost optimal. For stricter carbon dioxide limits, power generated from gas turbines is at first replaced by generation from increasing renewable capacities. Non-hydro storage capacities are only built for low-emission scenarios, in order to provide the necessary flexibility to meet peaks in the residual load.

\end{abstract}


\end{frontmatter}

\begin{table*}[!t]

\begin{framed}

\nomenclature[01]{$n$}{nodes (countries)}
\nomenclature[02]{$t$}{hours of the year}
\nomenclature[03]{$s$}{generation and storage technologies}
\nomenclature[04]{$\ell$}{inter-connectors}%
\nomenclature[05]{$c_{n,s}$}{fixed annualised generation and storage costs}
\nomenclature[06]{$c_\ell$}{fixed annualised line costs}
\nomenclature[07]{$o_{n,s}$}{variable generation costs}
\nomenclature[10]{$e_{s}$}{specific CO$_{2}$ emissions}%
\nomenclature[11]{$d_{n,t}$}{demand}
\nomenclature[12]{$g_{n,s,t}$}{generation and storage dispatch}
\nomenclature[13]{$\bar{g}_{n,s,t}$}{availability per unit of capacity}
\nomenclature[14]{$G_{n,s}$}{generation and storage capacity}
\nomenclature[15]{$G_{n,s}^{max}$}{maximum installable capacity}%
\nomenclature[20]{$f_{\ell,t}$}{power flow}
\nomenclature[21]{$F_{\ell}$}{transmission capacity}
\nomenclature[22]{$K_{n\ell}$}{incidence matrix}
\nomenclature[23]{$l_\ell$}{length of transmission line}
\nomenclature[27]{EU}{European Union}%
\nomenclature[27]{O\&M}{operation and maintenance}%
\nomenclature[27]{PV}{solar photovoltaic}
\nomenclature[27]{OCGT}{open-cycle gas turbines}
\nomenclature[27]{PHS}{pumped hydro storage}
\nomenclature[27]{H$_2$}{molecular hydrogen}%
\nomenclature[27]{HVDC}{high-voltage direct current}
\nomenclature[27]{HVAC}{high-voltage alternating current}
\nomenclature[27]{NTC}{net transfer capacity}%

\printnomenclature

\end{framed}

\end{table*}

%%%%%-----%%%%%-----%%%%%-----%%%%%------%%%%%
%%%%%-----%%%%%-----%%%%%-----%%%%%------%%%%%
\section{Introduction}
\label{sec:intro}
%%%%%-----%%%%%-----%%%%%-----%%%%%------%%%%%
%%%%%-----%%%%%-----%%%%%-----%%%%%------%%%%%
In order to meet the ambitious target of reducing carbon dioxide (CO$_2$) emissions in the European Union by 80\% to 95\% in 2050 compared to 1990 values, the electricity system has to undergo a fundamental transformation (see for instance the Energy Roadmap 2050 from the European Commission~\cite{eu2050}). Wind and solar power plants are already today both mature and cost-efficient technology options, which can be scaled up to act as the basis of a low-emission future power supply (see~\cite{IRENA2018a} for an analysis of the increasing  cost-competitiveness of renewable power generation technologies, and~\cite{IRENA2018b} for a discussion of cost-effective renewable energy options for all EU Member States). The challenges presented by the temporal fluctuations in these resources can be met with low-carbon technologies such as existing hydroelectricity power plants, or with storage options like batteries or hydrogen storage, which still have significant potential for further development. The benefit of the flexibility provided by storage has for instance been studied for a simplified model of a highly renewable  European electricity system in~\cite{Rasmussen2012}, or with a focus on pumped hydro storage and wind power generation for the Irish system in~\cite{Connolly2012}. In~\cite{budischak2013} the authors focus on least-cost combinations of renewable generation and storage for a large regional grid, whereas in~\cite{Cebulla2017} also the role of the spatial distribution and dispatch of storage capacities on continental scale in an European electricity system with 80$\%$ power production from variable renewable energies has been studied.

With respect to the spatial variability of weather-dependent renewable generation, large-scale power transmission capacities play a decisive role to provide a smoothing effect and to connect generation capacity at favourable distant locations with the load centres. The systemic advantage of aggregating variable renewable power generation over large distances has already been observed in the pioneering study by Czisch in~\cite{Czisch}. This  benefit of transmission has been confirmed in studies using more detailed models, which also take into account limitations and costs of the transmission infrastructure (see for instance~\cite{Schaber2} for a systematic study with regard to the renewable penetration levels and mixes, \cite{Gils2017} for additional detailed backup capacity optimisations, or \cite{schlachtberger_benefits_2017} for a techno\hyp{}economic optimisation study at high renewable shares.

Given the complexity of such a system, in particular with respect to the spatio-temporal patterns and correlations in the renewable generation and load time series, it is a difficult task to deduce a cost-efficient overall system layout from heuristic principles alone. As a consequence, computational models are a central element in the development of policy guidelines for the design of a future low-emission energy system (see~\cite{pfenninger_energy_2014} for a recent review). For each model, choices have to be made regarding the methodological scheme and the scope of its system representation (for instance the spatial and temporal scale, or the choice of energy sectors and technology options considered in the model). The resulting diversity of system models represents a challenge for the interpretation and comparison of the corresponding results, in particular if the underlying input data or modelling details are not publicly accessible. In this context, open energy modelling promises to provide a more transparent and comprehensible scientific approach (see the discussion in \cite{pfenninger_opening_2018} or~\cite{Pfenninger2017b}).

However, even for a single system model for which all input data and modelling details are transparent, the deeper understanding of the numerical results that is fundamental for robust policy advice can be hindered by the dependence of the results on the choice of the model parameters. In particular the details of the input data, the cost assumptions, and the constraints employed in the model all affect the properties of the resulting scenarios in a non-trivial way.

In this contribution these issues are addressed by studying in detail the influence of weather data, cost parameters and policy constraints on the properties of cost optimal scenarios of a future highly renewable electricity system taken from \cite{schlachtberger_benefits_2017}. It is shown that the total system costs are only weakly affected by the choice of the input weather data or by small changes in the capital costs. The optimisation landscape is flat in many directions, which allows system planners to choose between different near optimal system configurations without a significant increase in total costs. With respect to policy constraints, the investigation of a wide range of CO$_2$ emission limits helps to understand the mechanisms in the cost-efficient interplay of different technology options along the pathway towards a future low-emission electricity system.

The approach taken in the present contribution goes beyond the small variations of a few selected input parameters that are commonly used in the literature to test robustness. Here very large changes to input parameters are considered to try to understand fully what role each system component plays in the cost-optimal system.

In the literature, the consideration of different input assumptions is often limited to the focus of the respective study.  For instance in~\cite{schlachtberger_benefits_2017}, Schlachtberger et al. focus on the role of transmission expansion for the layout and cost structure of a low-emission European electricity system. The parameter variation is then mainly limited to the constraint for the total transmission capacity as the crucial parameter of the study, but not expanded  in a similar way to other model dimensions. Similarly, in~\cite{Schaber} the authors analyse the effects of grid extensions in a renewable Europe, with a focus on the influence of the composition of the power generation in the system. The screening of parameter ranges for the renewable penetration and the mix between solar and wind is the central line of the investigation. In other cases, a sensitivity analysis of modelling results is added to the presentation of a numerical study to assess the robustness of the numerical findings. Such an analysis usually addresses only a few key parameters, but is not central to the understanding of the workings of the system. For instance in~\cite{Gils2017}, the authors model a renewable-based European electricity system including storage options and concentrated solar power. Choosing different investment costs for storage, grid and backup technologies, they find that although the system composition and operation is highly dependent on these parameters, the overall system cost is only slightly affected.

This article starts with a review of the model and the data inputs in Sec.~\ref{sec:methods}. In Sec.~\ref{sec:input_data} results on the sensitivity to different samples of weather and load data and in Sec.~\ref{sec:cost} to changes in the cost assumptions are presented. The subsequent Sec.~\ref{sec:constraints} explores the role of policy constraints, such as limits on the expansion of onshore wind, or different CO$_2$ limits. Finally in Sec.~\ref{sec:discussion} the limitations of this study are discussed, before conclusions are drawn in Sec.~\ref{sec:conclusion}.

%%%%%-----%%%%%-----%%%%%-----%%%%%------%%%%%
%%%%%-----%%%%%-----%%%%%-----%%%%%------%%%%%
\section{Methods}
\label{sec:methods}
%%%%%-----%%%%%-----%%%%%-----%%%%%------%%%%%
%%%%%-----%%%%%-----%%%%%-----%%%%%------%%%%%
The following sections briefly review the underlying model and data as well as properties of some of the resulting scenarios of a future European electricity system discussed in~\cite{schlachtberger_benefits_2017}.
%%%%%-----%%%%%-----%%%%%-----%%%%%------%%%%%
\subsection{Model}
%%%%%-----%%%%%-----%%%%%-----%%%%%------%%%%%
The model presented in~\cite{schlachtberger_benefits_2017} applies a linear techno\hyp{}economic optimisation of total annual costs:
%-----%-----%------%
\begin{equation}
  \min_{G_{n,s},F_\ell,g_{n,s,t},f_{\ell,t}} \left( \sum_{n,s} c_{n,s} G_{n,s} + \sum_{n,s,t} o_{n,s} g_{n,s,t}  + \sum_{\ell} c_{\ell} F_{\ell} \right)~.
  \label{eq:objective}
\end{equation}
%-----%-----%------%
The indices $n$ label the nodes of the system, which represent with one node per country the EU-28 (as of 2018) without Cyprus and Malta, but including Norway, Switzerland, Serbia, and Bosnia and Herzegovina. The system costs are composed of fixed annualised costs $c_{n,s}$ for generation and storage capacity $G_{n,s}$, variable costs $o_{n,s}$ for generation and storage dispatch $g_{n,s,t}$, and fixed annualised costs $c_{l}$ for transmission capacity $F_{l}$. The indices~$s$ label the generation and storage technologies comprising onshore wind, offshore wind, solar PV, open cycle gas turbines (OCGT), hydrogen storage (electrolysis and fuel cells for conversion, steel tanks for storage), central batteries (lithium ion), pumped hydro storage, hydro reservoir, and run-of-river hydro generation.

The optimisation problem in Eq.~(\ref{eq:objective}) is subject to several constraints. Assuming an inelastic demand $d_{n,t}$ at node~$n$ and time~$t$, the instantaneous balancing of energy supply and demand translates into the power balance constraints
%-----%-----%------%
\begin{equation}
  \sum_{s} g_{n,s,t} - d_{n,t} = \sum_{\ell} K_{n\ell} f_{\ell,t}  \hspace{1cm}  \forall\, n,t~, \label{eq:balance}
\end{equation}
%-----%-----%------%
where $K_{n\ell}$ is the incidence matrix of the network. Here $f_{\ell,t}$ denotes the power flow at time $t$ on the line $\ell$ representing the cross-border transmission between the respective two  interconnected countries in the European electricity system. The power transmission is modelled as a transport model, given the increasingly controllable nature of international connections realised as point-to-point HVDC connections or using phase-shifting transformers in the case of HVAC connections. The absolute power flows have to respect the capacity limits
%-----%-----%------%
\begin{equation}
  |f_{\ell,t}| \leq F_{\ell} \hspace{1cm} \forall\,\ell,t~,
\end{equation}
%-----%-----%------%
with the system model determining the line capacities $F_{l}$ in its optimisation of total system costs in Eq.~(\ref{eq:objective}). In~\cite{schlachtberger_benefits_2017} a cap $\mathrm{CAP}_{LV}$ of the total transmission line capacities multiplied by their lengths was introduced,
%-----%-----%------%
\begin{equation}
\label{eq:cap_lv}
  \sum_{\ell} l_{\ell}\cdot F_{\ell}  \leq \mathrm{CAP}_{LV}~,
\end{equation}
%-----%-----%------%
with the influence of this constraint on system costs representing one focus of the study.

Further constraints apply to the dispatch $g_{n,s,t}$ of both conventional and renewable generators as well as the storage operation. While the conventional generators can be dispatched up to their nominal capacity $G_{n,s}$,
%-----%-----%------%
\begin{equation}
  0 \leq g_{n,s,t} \leq G_{n,s} \hspace{1cm} \forall\, n,s,t~,
\end{equation}
%-----%-----%------%
the potential renewable generation in each installed unit of the respective generators depends on the given weather conditions:
%-----%-----%------%
\begin{equation}
\label{eq:availability}
 0 \leq  g_{n,s,t} \leq \bar{g}_{n,s,t} \cdot G_{n,s} \hspace{1cm} \forall\, n,s,t~.
\end{equation}
%-----%-----%------%
Here the availability $\bar{g}_{n,s,t}$ times the capacity $G_{n,s}$ gives the maximum renewable energy generation according to the weather conditions at node $n$ and time $t$. Note that it is assumed that this potential renewable generation can always be curtailed, such that Eq.~(\ref{eq:availability}) describes the limits of the corresponding possible dispatch. In the following, the term `renewable generation' always refers to this dispatch after curtailment. Both the conventional and the renewable dispatch, as well as the installed capacity itself are a result of the optimisation in Eq.~(\ref{eq:objective}), with the maximum installed renewable capacity $G_{n,s}^{max}\geq G_{n,s}$ following an assessment of the geographic potentials. For the operation of the different storage technologies, the state-of-charge has to be consistent with the charging / discharging in each hour while respecting the storage capacity, with operational losses being taken into account by corresponding efficiency parameters. The storage capacities are assumed to be proportional to the power capacities, with the ratio $h_{s,max}$ representing the time in which a storage unit can be fully charged or discharged at maximum power ($h_{s,max}=$ 6h for batteries and pumped hydro storage as short-term storage options, and $h_{s,max}=$ 168h for hydrogen as long-term storage). Standing losses in the storage units are neglected. For the full storage equations consult~\cite{schlachtberger_benefits_2017}.

Given the political significance of greenhouse gas emission targets of the EU member states, the total amount of CO$_2$ emissions is a key characteristic of the system. Such an emission limit enters the model as a cap~$\mathrm{CAP}_{CO_{2}}$ on the total emissions, implemented using the specific emissions $e_{s}$ in CO$_{2}$-tonne-per-MWh of the fuel of generator type $s$ with efficiency $\eta_{s}$:
%-----%-----%------%
\begin{equation}
  \sum_{n,s,t} \frac{1}{\eta_{s}} g_{n,s,t}\cdot e_{s} \leq  \mathrm{CAP}_{CO_{2}}~.
  \label{eq:co2cap}
\end{equation}
%-----%-----%------%
The only generators in the model with a non-zero CO$_2$ emission are open-cycle gas turbines with $e_{s}=0.19$ tonne-CO$_2$/MWh$_{th}$, or equivalently $e_{s}\approx 490$ g-CO$_2$/kWh$_{el}$ (see Tab.~\ref{tab:costsassumptions}). For further details the reader is referred to~\cite{schlachtberger_benefits_2017}.

The model is implemented in the free software energy system framework PyPSA~\cite{PyPSA}.

%%%%%-----%%%%%-----%%%%%-----%%%%%------%%%%%
\subsection{Data}
\label{sec:data}
%%%%%-----%%%%%-----%%%%%-----%%%%%------%%%%%

All data underlying the model reviewed in the last section is extensively described in~\cite{schlachtberger_benefits_2017}. In the following this data is briefly outlined. The reader is referred to the aforementioned publication for further details.

%-----%-----%------%
%\begin{adjustbox}{angle=90}
%\Rotatebox{90}{%
\begin{table*}%
 %  \begin{sidewaystable}

\begin{threeparttable}
\caption{Input parameters based on 2030 value estimates from~\cite{schroeder2013} unless stated otherwise (Table taken from~\cite{schlachtberger_benefits_2017}).}
\label{tab:costsassumptions}
\begin{tabularx}{\textwidth}{lrrrrrrr}
\toprule

Technology & \multicolumn{1}{l}{investment} & \multicolumn{1}{l}{fixed O\&M} & \multicolumn{1}{l}{variable} & \multicolumn{1}{l}{lifetime} & \multicolumn{1}{l}{efficiency} & \multicolumn{1}{l}{capital cost per} & \multicolumn{1}{l}{$h_{max}$} \\
 & \multicolumn{1}{l}{(\euro/kW)} & \multicolumn{1}{l}{cost} & \multicolumn{1}{l}{cost} & \multicolumn{1}{l}{(years)} & \multicolumn{1}{l}{(fraction)} & \multicolumn{1}{l}{energy storage} & \multicolumn{1}{l}{(h)} \\
 & \multicolumn{1}{l}{} & \multicolumn{1}{l}{(\euro/kW/year)} & \multicolumn{1}{l}{(\euro/MWh)} & \multicolumn{1}{l}{} & \multicolumn{1}{l}{} & \multicolumn{1}{l}{(\euro/kWh)} &  \\

\midrule

onshore wind & 1182 & 35 & 0.015\tnote{a} & 25 & 1 &  &  \\
offshore wind & 2506 & 80 & 0.02\tnote{a} & 25 & 1 &  &  \\
solar PV & 600 & 25 & 0.01\tnote{a} & 25 & 1 &  &  \\
OCGT\tnote{b} & 400 & 15 & 58.4\tnote{c} & 30 & 0.39 &  &  \\
hydrogen storage\tnote{d} & 555 & 9.2 & 0 & 20 & \multicolumn{1}{r}{$0.75\cdot 0.58$\tnote{e}} & 8.4 & 168 \\
central battery (LiTi)\tnote{d} & 310 & 9.3 & 0 & 20 & \multicolumn{1}{r}{$0.9 \cdot 0.9$\tnote{e}} & 144.6 & 6 \\
transmission\tnote{f} & {400 \euro /MWkm} & 2\% & 0 & 40 & 1 &  &  \\
PHS & 2000\tnote{g} & 20 & 0 & 80 & 0.75 & N/A\tnote{g} & 6 \\
hydro reservoir & 2000\tnote{g} & 20 & 0 & 80 & 0.9 & N/A\tnote{g} & fixed\tnote{h} \\
run-of-river & 3000\tnote{g} & 60 & 0 & 80 & 0.9 &  &  \\
\bottomrule
\end{tabularx}

\begin{tablenotes}
\item [a] The order of curtailment is determined by assuming small variable costs for renewables.
\item [b] Open-cycle gas turbines have a CO$_2$ emission intensity of 0.19 CO$_2$-tonne/MWh$_{th}$.
\item [c] This includes fuel costs of 21.6 \euro/MWh$_{th}$.
\item [d] Taken from~\cite{budischak2013}.
\item [e] The storage round-trip efficiency consists of charging and discharging efficiencies $\eta_1 \cdot \eta_2$.
\item [f] Taken from~\cite{Hagspiel}. Costs for converter pairs of 150000\euro/MW and an (n-1) security factor of 1.5 are taken into account for the transmission cost assumptions (see~\cite{schlachtberger_benefits_2017} for details).
\item [g] The installed facilities are not expanded in this model and are considered to be amortised. %
\item [h] Determined by size of existing energy storage, taken from~\cite{ENTSOEinstalledcapas,kies2016}.

\end{tablenotes}
\end{threeparttable}
 %  \end{sidewaystable}

\end{table*}
%\end{adjustbox}
%-----%-----%------%

For each scenario the model is run over a time series comprising one or multiple years with hourly resolution. The load in each country is given by the time series provided by the transmission system operators in~\cite{entsoe_load}. The renewable generation from onshore wind, offshore wind, and solar photovoltaic (PV) power generation are derived from historic weather data with a temporal hourly and a 40 $\times$ 40 km$^2$ spatial resolution. Using models for the renewable generation infrastructure, this weather data is converted into a potential wind/solar generation time series for each cell (see~\cite{Heide2010} and~\cite{Andresen2015} for a discussion of this procedure). By applying suitable capacity layouts, the resulting spatially-detailed potential generation data is then aggregated into generation time series on country level. These capacity layouts take into consideration the spatial distribution of the renewable resource quality inside the countries, as well as the geographical potentials considering the land use type, nature reserves, restricted areas, and likely public acceptance (data sources include the CORINE land cover~\cite{corine2006} and Natura 2000~\cite{natura2000} databases, see also \cite{Scholz} for a more general discussion of renewable generation potentials). Hydroelectricity generation comprises reservoir hydro and run-of-river power plants according to the currently installed capacity, with the hourly generation following inflow time series on country level (for data about installed capacity and a description of the methods to derive hydro power generation time series see~\cite{kies2016,pfluger2011,ENTSOEinstalledcapas,dee2011}). As a conventional backup system the model can build flexible open-cycle gas turbines (OCGT) (data for generation costs and emissions are taken from~\cite{schroeder2013}), whose global annual energy generation is restricted by the EU CO$_2$ emission limit represented by the cap $\mathrm{CAP}_{CO_{2}}$. As possible storage technologies, the model includes pumped hydro storage (PHS), central batteries, and hydrogen (H$_{2}$) storage, with the hydro storage assumed to correspond to the currently installed capacities (the corresponding data is taken from~\cite{budischak2013} and~\cite{kies2016}). The cost assumptions regarding all generation and storage technologies are summarised in Tab.~\ref{tab:costsassumptions}. A discount rate of 7$\%$ is assumed in order to convert overnight capital costs into net present costs.

%%%%%-----%%%%%-----%%%%%-----%%%%%------%%%%%
\subsection{The base scenarios}
%%%%%-----%%%%%-----%%%%%-----%%%%%------%%%%%

As the base scenarios for the present study three configurations with different interconnecting line volumes reported in~\cite{schlachtberger_benefits_2017} are chosen. Building on consumption and generation time series for the year 2011, the system optimisation according to Eq.~(\ref{eq:objective}) is performed for a CO$_2$ cap corresponding to a reduction of 95$\%$ of European CO$_{2}$ emissions compared to 1990. Three different levels of total transmission capacity interpolate between a locally-balanced, storage-dominated system and a continental-scale, network-dominated system: The zero transmission scenario neglects any transmission capacity between the European countries, resulting in independently optimised systems for each of the European countries, with an average system cost of 84.1 \euro/MWh. The optimal transmission scenario does not invoke any exogenous limit to the total transmission capacity. This allows a reduction of total systems costs to 64.8 \euro/MWh, with a total line volume of 286 TWkm or approximately nine times the sum of todays net transfer capacities (NTCs) of 31 TWkm. The compromise transmission scenario restricts the total line volume in Eq.~(\ref{eq:cap_lv}) to an intermediate value of four times todays total capacities, $\mathrm{CAP}_{LV}$=125 TWkm, which already locks in a reduction of system costs to 67.5 \euro/MWh. The main characteristics of these three base scenarios are reviewed in Tab.~\ref{tab:results_costs}; for further details consult~\cite{schlachtberger_benefits_2017}.

%-----%-----%------%
\begin{table}
\centering
\caption{Optimised average system costs in [\euro/MWh] for the allowed total interconnecting line volume of the zero, compromise, and optimal grid scenarios. Also given is the total line volume (adapted from~~\cite{schlachtberger_benefits_2017})}
\begin{tabularx}{\columnwidth}{lrrr}
  \toprule
 {\bf Scenario}  & {\bf Zero}  &  {\bf Comp.} & {\bf Opt.} \\
  \midrule
Line vol. [TWkm]   &  0.0   &  125.0 &  285.70 \\
\midrule
battery storage    &     9.9  &     4.5 &     1.7 \\
hydrogen storage   &     8.1 &     3.4 &     3.1 \\
gas                &     4.6  &     4.1 &     4.5 \\
solar              &    26.1 &    14.7 &     9.4 \\
onshore wind       &    22.3 &    23.4 &    28.6 \\
offshore wind      &    10.8  &    11.4 &     7.5 \\
transmission lines &     0.0  &     3.6 &     7.6 \\
PHS                &     0.3  &     0.3 &     0.3 \\
run-of-river       &     1.4 &     1.4 &     1.4 \\
reservoir hydro    &     0.8  &     0.8 &     0.8 \\
\midrule
Total cost         &    84.1  &    67.5 &    64.8 \\
\bottomrule
\end{tabularx}
\label{tab:results_costs}
\end{table}
%-----%-----%------%

%%%%%-----%%%%%-----%%%%%-----%%%%%------%%%%%
%%%%%-----%%%%%-----%%%%%-----%%%%%------%%%%%
\section{Results I: Sensitivity to the input weather and load time series}
\label{sec:input_data}
%%%%%-----%%%%%-----%%%%%-----%%%%%------%%%%%
The base scenarios presented in~\cite{schlachtberger_benefits_2017} build on time series for renewable generation potential and consumption with hourly resolution for the year 2011. Previous investigations in~\cite{Pfenninger2017} and~\cite{Kotzur2018} show that the outcomes of energy system models depend to some degree on the sampling and time resolution of the input data as well as the data reduction method. Similar observations have been made in~\cite{Hartel2017} in the context of the role of sampling and clustering techniques for offshore grid expansion planning. Given the decisive role of weather dependent generation from wind and solar in relation to the consumption patterns for a low emission electricity system, it is expected that the optimised layouts and the associated cost structures studied in~\cite{schlachtberger_benefits_2017} depend on the choice of the input data. In this section the sensitivity of system costs to changes in the load and renewable generation time series is explored.

\paragraph{Different single historical weather years}
First, the single year optimisations of the base scenarios (2011) are repeated for the weather and load data for the years 2012 to 2014. This limited time frame is not further expanded due to the restricted availability of consistent load and hydro inflow datasets.

\begin{figure}
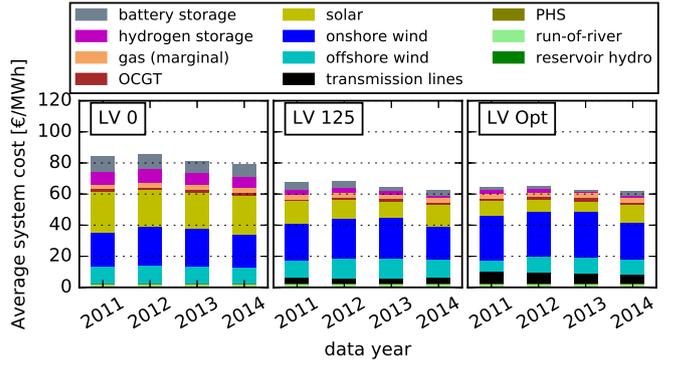
%[htbp]
\centering
\includegraphics[width=\linewidth]{{{diw2030_solar1_7_year/avgcosts_wHydroFOM_LVall_diw2030_solar1_7_year}}}
\caption{Composition of the average total system costs per unit of generated energy for the four different weather and demand input years 2011 to 2014 (left to right bars) for the zero, compromise, and optimal (left to right panels) transmission grid scenarios.}
\label{fig:year_cost}
\end{figure}

Fig.~\ref{fig:year_cost} shows that the optimal system configuration depends to some extent on the simulated year, with the average system costs for the compromise grid volume ranging from 62.6 to 67.9 \euro/MWh.

%-----%-----%------%
%-----%-----%------%%-----%-----%------%

\begin{figure}
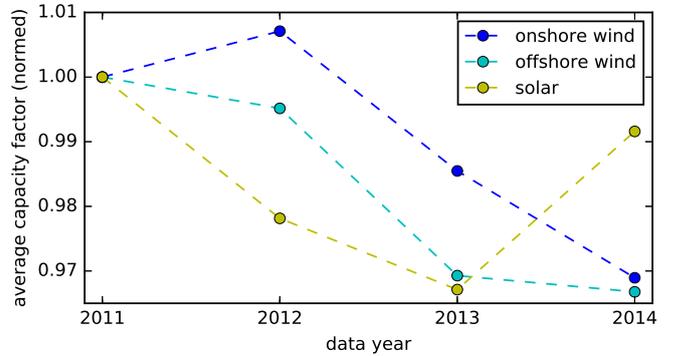

\centering
\includegraphics[width=\linewidth]{{{diw2030_solar1_7_year/avgcapafactor_normed_diw2030_solar1_7_year}}}
\caption{
Average capacity factor $\mathrm{cf}_{s}$ of onshore wind (blue), offshore wind (cyan), and solar (yellow) for different weather years, normalised to $\mathrm{cf}_{s}= 0.233$, $0.485$, $0.128$ for offshore wind, onshore wind, and solar obtained for the data year 2011. A fixed capacity layout given by the installation potential $G_{n,s}^{max}$ has been assumed for all years. The data points are marked by circles, with the dashed lines drawn as visual aids. Note that the y-axis range is limited to [0.965,1.01] to show the relative changes.}
\label{fig:year_capa}
\end{figure}

In order to assess the year-dependent resource quality of renewable power generation, the annual average capacity factor $\mathrm{cf}_{s}$ is defined as
%-----%-----%------%
\begin{equation}
\mathrm{cf}_{s} = \frac{\sum_{n} \langle \bar{g}_{n,s,t} \rangle_{t} \cdot G_{n,s}^{max} }{\sum_{n} G_{n,s}^{max}}~,
\end{equation}
%-----%-----%------%
where the temporal average $\langle \cdot \rangle_{t}$ is taken over all hours of the year under consideration. This measures the average amount of energy that can be produced per unit of installed capacity. To allow for a clearer comparison of the weather conditions, a fixed capacity layout is chosen, given by the installation potential $G_{n,s}^{max}$ for all years rather than the year-dependent layout $G_{n,s}$ resulting from the system optimisation. Fig.~\ref{fig:year_capa} displays the relative changes of the values compared to the capacity factor $\mathrm{cf}_{s}= 0.485, 0.233, 0.128$ for offshore wind, onshore wind, and solar obtained for the base scenarios data year 2011. These relative changes are smaller than 3.5\% for the considered weather years.

%-----%-----%------%
%-----%-----%------%

The composition of the system in terms of shares of onshore wind and solar installations is largely determined by the respective capacity factors. For the weather years with capacity factor lower than in 2011, the installed capacities are also proportionately lower, and vice versa (compare Figs.~\ref{fig:year_cost} and~\ref{fig:year_capa}). The size of offshore wind installations does not follow the same trend but remains relatively stable. This can be explained by the fact that offshore wind is only built in a few countries and that the average capacity factor is much higher and therefore the benefit of a smoother generation profile is more important than a slightly lower capacity factor.

\begin{figure}
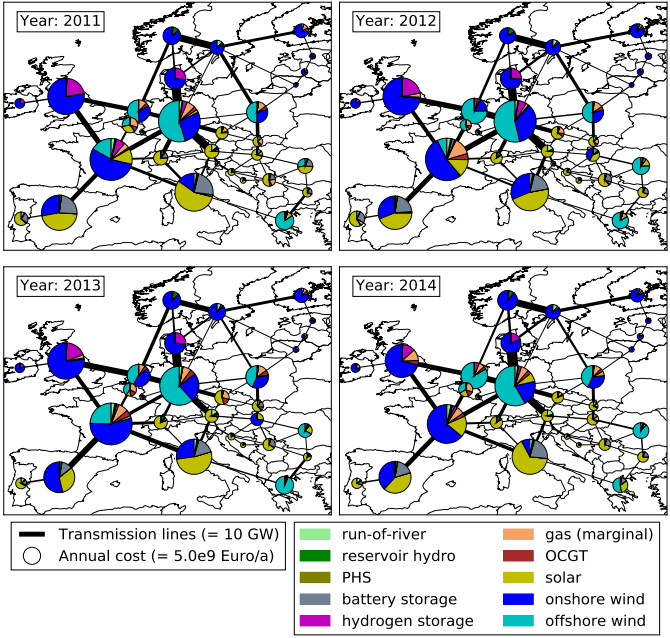

\centering
\includegraphics[width=\linewidth]{{{diw2030_solar1_7_year/network-euro-pie-costs_hydro-diw2030_solar1_7_year-LV0.25-combined-2011-2012-2013-2014}}}
\caption{Map of average annual system costs per country in the compromise transmission scenario for the four individual years of weather and demand data 2011 to 2014 (left to right, top to bottom).
The area of the circles is proportional to the total costs per country. The modelled international transmission lines are shown as black lines with width proportional to their optimised net transfer capacity (see~\cite{schlachtberger_benefits_2017} for details).}
\label{fig:year_cost_map}
\end{figure}

The regional distribution of the generators in the compromise transmission scenario is shown in Fig.~\ref{fig:year_cost_map} and follows the expected trends. In the data year 2013, for which the solar capacity factor is lowest, the countries that are usually solar dominated such as Spain and Italy see cost optima with significant shares of onshore wind, while in 2014, when the total solar share is relatively high, there are solar installations as far north as Poland. The transmission capacity has to be redistributed only to a small extent to adapt to the slightly different generation capacity layouts.

%-----%-----%------%
%-----%-----%------%

Note the higher total system costs in the data year 2012 compared to 2011 (1.6\% / 0.65\% / 0.67\% for the scenario without / with moderate / with optimal transmission), despite the higher onshore wind and only marginally lower offshore wind capacity factor. This relation indicates that weather conditions corresponding to a higher capacity factor alone do not necessarily reduce the total system costs. Spatio-temporal correlations in the combined load and generation time series are also important to determine the cost optimum.

\begin{figure}
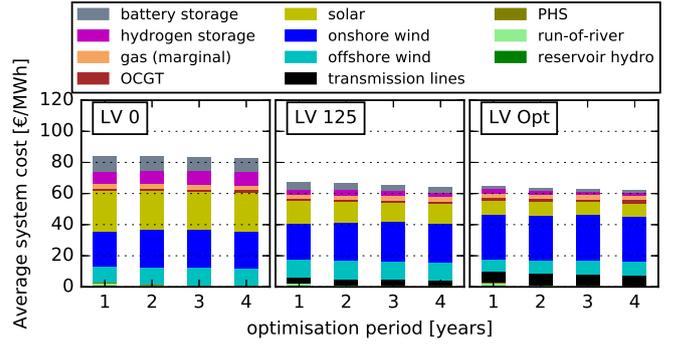
%[htbp]
\centering
\includegraphics[width=\linewidth]{{{diw2030_solar1_7_multiyear/avgcosts_wHydroFOM_LVall_diw2030_solar1_7_multiyear}}}
\caption{Composition of the average system costs per unit of generated energy for different numbers of simulated weather and consumption years starting from 2011 for the zero, compromise, and optimal (left to right panels) transmission grid scenarios.
The color code indicates the contributions of different technologies.}
\label{fig:multiyear_cost}
\end{figure}

%-----%-----%------%%-----%-----%------%
\paragraph{Multi-year optimisation}
%-----%-----%------%%-----%-----%------%
Instead of modelling the system for a single year of weather and load data, it can also be optimised for a longer period of continuous input data. This inclusion of a larger dataset might result in a solution that is less sensitive to changes in the input parameters due to its adaption to a potentially larger spectrum of generation and demand situations.
Starting from the weather and load year 2011 considered in the base scenarios, a simultaneous optimisation over one to four years of continuous data is considered. The limitation to four years of data is founded in the restricted availability of consistent datasets and in the considerable computational demands  of running multi-year optimisations (for a discussion of the software implementation of the model see~\cite{schlachtberger_benefits_2017}). Fig.~\ref{fig:multiyear_cost} shows that for the compromise grid scenario the average total system costs $67.1 \pm 0.7$ \euro/MWh deviate very little from the value $67.5$ \euro/MWh of the base scenario. This result is also close to the maximum cost of the single year simulations (2012: 67.8 \euro/MWh), indicating that the system capacity is set by a few extreme events over the whole period. The maximum cost difference between the four year and the single year simulations is 3.5 \euro/MWh, a 5.5\% increase from the value of the weather year 2014. In the range of the time span considered in this study, this is a lower bound for the uncertainty of a single year optimisation compared to longer-term modelling.

%-----%-----%------%
%-----%-----%------%

\begin{figure}
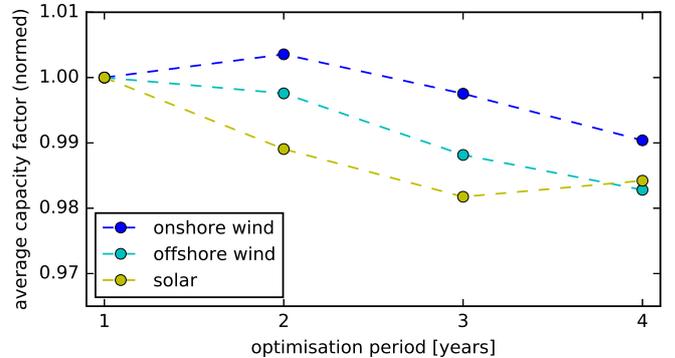

\centering
\includegraphics[width=\linewidth]{{{diw2030_solar1_7_multiyear/avgcapafactor_normed_diw2030_solar1_7_multiyear}}}
\caption{Same as Fig.~\ref{fig:year_capa}, but for capacity factor $\mathrm{cf}_{s}$ averaged over different numbers of years starting from 2011.}
\label{fig:multiyear_capa}
\end{figure}

Similar to the single year optimisations, the shares of onshore wind and solar installations in the multi-year models are proportional to the corresponding average capacity factors. However, the changes in both quantities are significantly smaller, as indicated in Fig.~\ref{fig:multiyear_capa}.

%-----%-----%------%
%-----%-----%------%

%-----%-----%------%%-----%-----%------%
\paragraph{3h sampling}
%-----%-----%------%%-----%-----%------%
Some weather data sets, e.g.~\cite{jacob2014}, provide data only in a lower temporal frequency of 3h. Computational restrictions might also prevent an optimisation at a higher temporal resolution, in particular if a very long time frame is considered. At lower temporal resolution, the number of optimisation variables and therefore the memory requirements are proportionately lower and also the computational solving time potentially decreases by a similar magnitude. Both can be limiting factors.

\begin{figure}
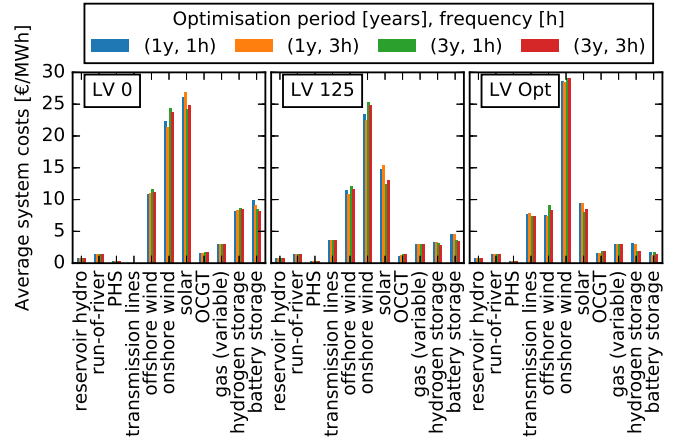
%[htbp]
\centering
\includegraphics[width=\linewidth]{{{diw2030_solar1_7_sub/avgcosts_pertype_LVall_diw2030_solar1_7_sub}}}
\caption{Average system cost of all modelled technologies over total consumption for all four combinations of 1 and 3 simulated years starting from 2011 at 1-hour and 3-hour time resolution (left to right bars) for the zero, compromise, and optimal (left to right panels) transmission grid scenarios.
The time resolution was decreased by taking 3-hour means of the hourly values of the demand and potential renewable generation time series.}
\label{fig:3h_cost}
\end{figure}

In the following a reduced time resolution by taking 3-hour means of the hourly values of the demand and potential renewable generation time series is considered. The system is then optimised both for the base scenarios year 2011 and the three-year period 2011 to 2013, such that both data sets consider the same number of time steps. Fig.~\ref{fig:3h_cost} shows that the share of solar power slightly increases and replaces wind if the sampling frequency is reduced. This is due to the much stronger fluctuations of solar generation on an hourly timescale compared to wind power generation. The temporal averaging implicitly simulates the smoothing effects of a short term storage, which becomes also apparent in the reduced battery capacity. This effect is not as pronounced in the wind time-series as their dominant fluctuations occur on larger timescales. This indicates that models with time resolution less frequent than one hour tend to overestimate the effectiveness of solar generation and therefore underestimate battery and wind generation requirements. Similar results have been seen in the literature when increasing the sampling resolution higher than one hour, for example to 5 minute intervals: only minor changes to system costs are seen \cite{DEANE2014152}, but flexibility may be undervalued \cite{6345631,ODwyer2015}; a more general discussion of time resolution in energy system models can be found in \cite{burdenresponse}. However, in the scenarios discussed here, the overall effects are small and are in fact outweighed by the fluctuations due to modelling a different period of weather and demand data. Therefore a reduction to 3-hour sampling seems reasonable for long-term planning studies, particularly if it enables more weather years to be considered.

%-----%-----%------%
%-----%-----%------%

%%%%%-----%%%%%-----%%%%%-----%%%%%------%%%%%
\section{Results II: Sensitivity to cost assumptions}
\label{sec:cost}
%%%%%-----%%%%%-----%%%%%-----%%%%%------%%%%%

Estimates for the future development of costs for technologies that are yet to undergo very large-scale deployment are intrinsically uncertain. In particular solar PV and storage costs could potentially drop significantly over the next decades (see \cite{schroeder2013} for an overview of prospective electricity generation costs until 2050, and in particular~\cite{etip} for a discussion of the cost development of solar PV). Since cost assumptions are a crucial input parameter for the optimisation approach in energy system modelling, a comprehensive system analysis has to quantify the sensitivity to cost assumptions in the respective scenarios. This important point is assessed here by varying the capital investment and fixed operation and maintenance cost assumptions for one technology at a time over a large range while keeping all others at their base scenario value.

%-----%-----%------%%-----%-----%------%
\paragraph{Solar capital costs}
%-----%-----%------%%-----%-----%------%
In the base scenario, investment costs for solar PV are assumed to be 600 \euro~per kW of installed capacity. This value is already an extrapolation to the year 2030 and less than half of the cost in 2010, but could be reduced by another 30\% until 2050 \cite{schroeder2013}. In \cite{etip} costs for utility-scale PV are foreseen to drop to around 240~\euro/kW by 2050, which is a drop of 60\% compared to the base value.

\begin{figure}
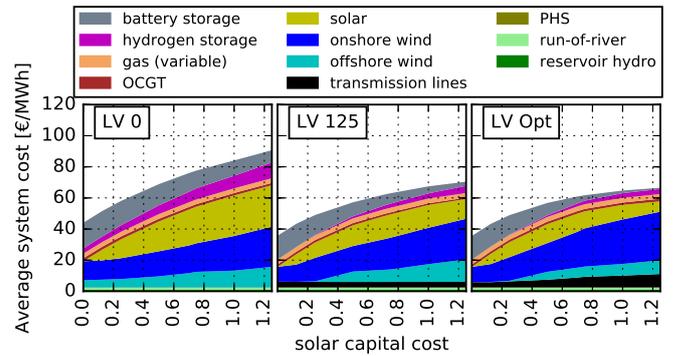
%[htbp]
\centering
\includegraphics[width=\linewidth]{{{diw2030_solar1_7_solarcapital/avgcosts_wHydroFOM_LVall_diw2030_solar1_7_solarcapital}}}
\caption{Composition of the average total system costs per unit of generated energy as a function of the fraction of the base solar PV capital cost assumption for the zero, compromise, and optimal (left to right panels) transmission grid scenarios.}
\label{fig:solarcapital_cost}
\end{figure}

If the solar capital costs are assumed to be 70\% of the base value, or 420 \euro/kW, for the compromise grid scenario the average total system costs are reduced by 9.6\% from 67.5 to 61 \euro/MWh, as shown in Fig.~\ref{fig:solarcapital_cost}.
This reduction of 6.5 \euro/MWh is slightly larger than the direct cost decrease of 4.4 \euro/MWh that would occur if only the costs, but not the installed capacity of solar PV are changed. This implies an effective additional benefit to the system due to the shift to a higher solar share from 0.61 TW to 0.87 TW of 2.1~\euro/MWh, or 3.1\%~of the total costs when reducing solar costs by~30\%.

%-----%-----%------%
%-----%-----%------%

\begin{figure}
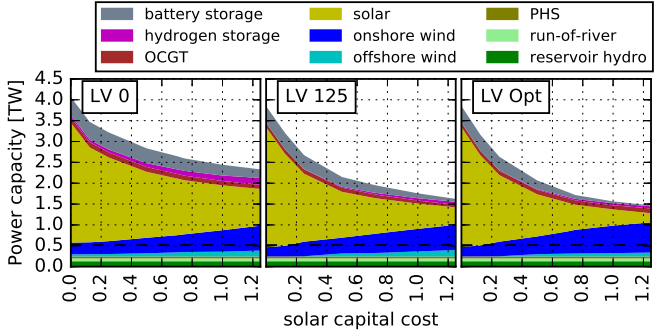
%[htbp]
\centering
\includegraphics[width=\linewidth]{{{diw2030_solar1_7_solarcapital/powercapa_wHydroFOM_LVall_diw2030_solar1_7_solarcapital}}}
\caption{Composition of the total installed power capacity of generators and storage units as a function of the fraction of the base solar PV capital cost assumption for the zero, compromise, and optimal (left to right panels) transmission grid scenarios. The peak of the total demand of 517 GW is marked as a horizontal dashed black line.}
\label{fig:solarcapital_power}
\end{figure}

Even though the total system costs are sensitive to solar cost assumptions, most changes are linear with moderate slopes down to 50\%--70\% of the solar cost. With decreasing solar costs, the installed capacity, generated energy, and curtailment of wind decrease. It is replaced by additional solar and battery capacity. Additionally, for increasing transmission capacity the power capacity of H$_2$ storage is almost completely replaced by OCGT capacity (see Fig.~\ref{fig:solarcapital_power}). This indicates that balancing the demand locally becomes a more dominant solution compared to spatial and long-term smoothing of wind generation, especially in countries with good solar resources. This is also reflected in the transmission volume in the optimal grid scenario, which decreases significantly, but in this range still stays above 125 TWkm, the volume of the compromise grid. A further indication that lower solar costs lead to decentral solutions is the difference between the system costs of the LV 0 and LV Opt scenarios: with the base solar cost, LV 0 is 30\% more expensive than LV Opt, but with 70\% lower solar cost, LV 0 is only 18\% more expensive.

%-----%-----%------%
%-----%-----%------%

\begin{figure}
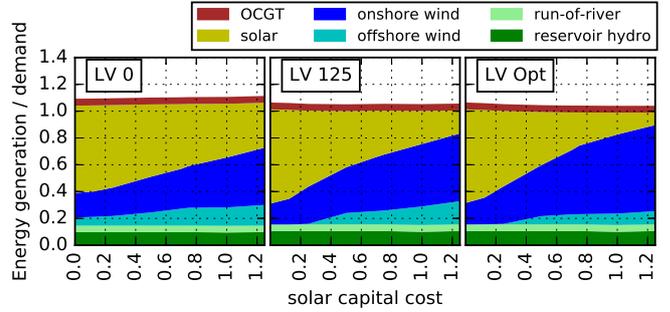
%[htbp]
\centering
\includegraphics[width=\linewidth]{{{diw2030_solar1_7_solarcapital/energygen_wHydroFOM_LVall_diw2030_solar1_7_solarcapital}}}
\caption{Composition of the total generated energy in units of the total demand as a function of the fraction of the base solar PV capital cost assumption for the zero, compromise, and optimal (left to right panels) transmission grid scenarios. Energy generation above the demand is caused by losses from storage use. The amount of curtailed energy is not shown.}
\label{fig:solarcapital_energy}
\end{figure}

In the extreme case that solar generation capacity becomes very cheap compared to the other technologies, even for the optimal grid scenario there is an upper limit of 70\% of the consumed energy that can be provided by solar alone, as shown in Fig.~\ref{fig:solarcapital_energy}. However, such a system set-up features huge solar power capacities of up to 2.9 TW, which corresponds to 5.6 times the peak of the total demand. At the same time, a significant amount of solar energy of up to 30\% of the yearly consumption has to be curtailed, leading to an inefficient system with a low effective capacity factor. At 40\% of the solar cost assumption relative to the base scenario, this technology provides 50\% of the energy, with a power capacity of three times the peak of total demand and  only a small amount of curtailment. This scenario thus displays a comparatively high efficiency.

%-----%-----%------%
%-----%-----%------%

\begin{figure}
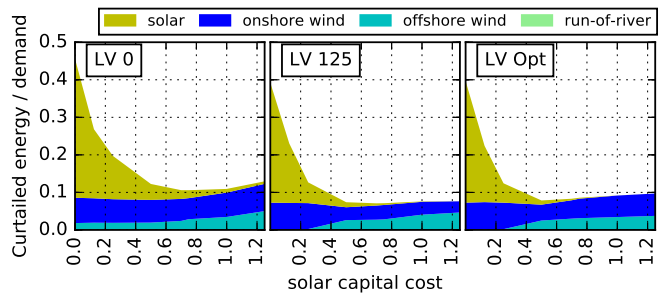
%[htbp]
\centering
\includegraphics[width=\linewidth]{{{diw2030_solar1_7_solarcapital/curtailedEnergy_wHydroFOM_LVall_diw2030_solar1_7_solarcapital}}}
\caption{Composition of the total curtailed energy in units of the total demand as a function of the fraction of the base solar PV capital cost assumption for the zero, compromise, and optimal (left to right panels) transmission grid scenarios. The order in which renewables are curtailed is: first offshore wind, then onshore wind, then solar.}
\label{fig:solarcapital_curtail}
\end{figure}

Fig.~\ref{fig:solarcapital_curtail} shows that the amount of curtailed energy remains relatively constant with decreasing solar capital costs at around 8\% to 12\% of the demand until solar costs are roughly cut in half. Even lower costs lead to the installation of significant overcapacities of solar generation and therefore a strong increase in curtailed energy of up to 40\% of the demand. In reality, this overproduction from solar could be used in other energy sectors such as transport and heating, before it is curtailed.

%-----%-----%------%
%-----%-----%------%

%-----%-----%------%%-----%-----%------%
\paragraph{Onshore and offshore wind capital cost}
%-----%-----%------%%-----%-----%------%
Wind turbines are a mature technology option and have already been deployed on a large scale in recent years, even though there is still significant additional installation potential in Europe. In particular offshore wind turbines are expected to undergo further technological development that will lead to significant cost reductions. In~\cite{schroeder2013} a decrease to 1075 and 2093 \euro/kW by 2050 for onshore wind and offshore wind, respectively, is assumed, which corresponds to a 9.1\% and 16.5\% reduction from the 2030 costs.

%-----%-----%------%
\begin{figure}
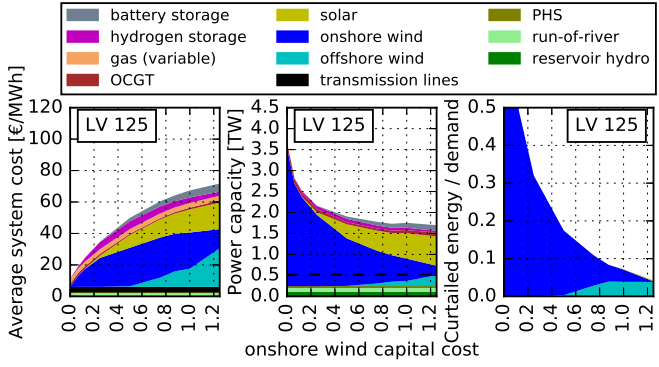
%[htbp]
\centering
\includegraphics[width=\linewidth]{{{opteu_onwindcapital/combined_cost_power_curtail_wHydroFOM_LV0.25_opteu_onwindcapital}}}
\caption{Left: Composition of the average total system costs per unit of generated energy as a function of the fraction of the base onshore wind capital cost assumption. In this figure, all panels are for the compromise transmission grid scenario.
Middle: Composition of the total installed power capacity of generators and storage units as a function of the fraction of the base onshore wind capital cost assumption. The peak of the total demand is marked as horizontal dashed black line.
Right: Composition of the total curtailed energy in units of the total demand as a function of the fraction of the base onshore wind capital cost assumption. The curtailment at zero cost corresponds to 1.3 times the total demand.}
\label{fig:on_cost_pow_curt}
\end{figure}
%-----%-----%------%

The optimisation results are already sensitive to moderate changes of the wind cost assumptions. Fig.~\ref{fig:on_cost_pow_curt} shows that for the compromise grid a 25\% reduction of the onshore wind cost decreases the total system costs by 10.4\%. While the onshore wind power capacity increases by 72.9\%, half of the offshore wind and a third of the solar installations are no longer required. This higher share of onshore wind generators also leads to a 51.5\% increase of the curtailed energy to 10.9\% of the total demand.
Assuming a less plausible more dramatic decrease of the onshore wind costs, these trends would continue until no offshore wind is built at 50\% cost reduction, and almost no solar at 25\% of the base scenarios cost assumption. The replacement of the remaining solar capacities is possible with excessive onshore wind power capacity, but would lead to a very large amount of curtailed energy of up to 1.3 times the annual demand. Considering the contrary scenario for a 25\% increase of onshore wind costs, up to 42.2\% of its installations would be replaced by offshore and solar capacity.

\begin{figure}
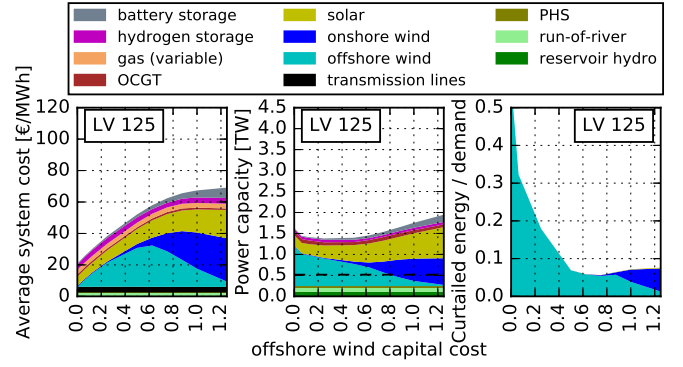
%[htbp]
\centering
\includegraphics[width=\linewidth]{{{opteu_offwindcapital/combined_cost_power_curtail_wHydroFOM_LV0.25_opteu_offwindcapital}}}
\caption{Same as Fig.~\ref{fig:on_cost_pow_curt} but as a function of the fraction of the base offshore wind capital cost assumption. The curtailment at zero cost corresponds to 0.57 times the total demand.}
\label{fig:off_cost_pow_curt}
\end{figure}

The total system costs are less sensitive to the offshore wind cost assumptions, but still decrease by 7.4\% for a 25\% lower offshore wind cost for the compromise grid, as shown in Fig.~\ref{fig:off_cost_pow_curt}.
Reducing offshore wind costs leads to a linear increase of its power capacity. It replaces dominantly onshore wind while the solar capacity remains relatively stable and is only slightly decreased.
This leads to a significant decrease of the total power capacity by up to 21.3\% compared to the base case at a large range of 50\% to at least 10\% of the assumed cost. At the same time, the amount of curtailed energy remains at or even below the initial value until the offshore wind cost assumptions are halved.

%-----%-----%------%
%-----%-----%------%

In the limit of very small offshore wind costs, the share of onshore wind is negligible, but solar can still contribute to a cost-optimal system with 241 GW or 14.8\% of the total power capacity. This suggests a positive correlation due to a different spatial distribution and the different generation profile of solar. These influences contribute to mitigate line congestions during hours of high demand if large amounts of power have to be transported from remote offshore wind installations.

\begin{figure}
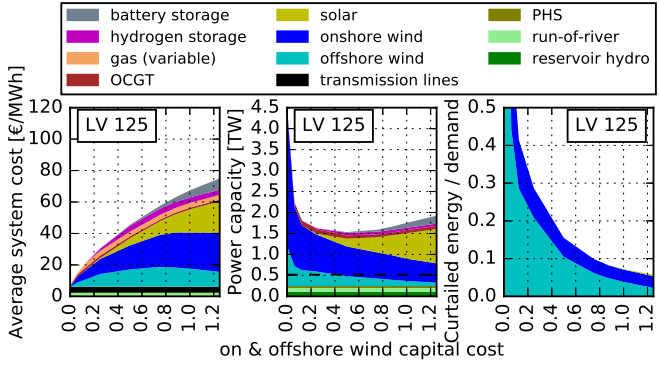
%[htbp]
\centering
\includegraphics[width=\linewidth]{{{opteu_onoffcapital/combined_cost_power_curtail_wHydroFOM_LV0.25_opteu_onoffcapital}}}
\caption{Same as Fig.~\ref{fig:on_cost_pow_curt} but as a function of the fraction of both the base onshore wind and the offshore wind capital cost assumption. The curtailment at zero cost corresponds to 2.6 times the total demand.}
\label{fig:onoff_cost_pow_curt}
\end{figure}

Fig.~\ref{fig:onoff_cost_pow_curt} shows that the system costs decrease even faster than in the previous cases if both the onshore and offshore wind cost assumptions are reduced simultaneously. As in the offshore wind case, the total installed power capacity decreases as the offshore wind installations are expanded and can provide more energy to the system, but a significant amount of solar power is now replaced by an increasing share of onshore wind. For very small cost assumptions, the results are similar to the onshore wind case with additional offshore wind installations that come with even larger amounts of curtailed energy.

Note that even for vanishing solar capital costs, the share of wind power generation in cost optimal configurations does not drop to zero (see Figs.~\ref{fig:solarcapital_cost} -~\ref{fig:solarcapital_curtail}). In contrast, already for moderate transmission expansion for very low onshore capital costs the optimisation leads to a scenario without solar power generation capacity. The fact that it is possible to set up an efficient system without solar, but not without wind power generation, shows the systemic value in the European system of the latter, which provides energy at day and night and is seasonally aligned with the peak demand. In contrast, meeting the winter peak with solar energy alone would require significant seasonal storage capacities. Synoptic variations of wind power as weather systems pass over Europe can be smoothed by the expansion of the continent-scale transmission system. This result was already indicated in an early study~by Czisch in~\cite{Czisch}, which demonstrated a cost-effective wind-dominated system for Europe, North Africa and the Middle East. The importance of the more system-friendly production of wind, particularly when integrated over large areas, also explains why the total system costs are more sensitive to changes in onshore wind capital costs than those of solar. Offshore wind is only available at specific locations, so profits less from the smoothing effect.

%-----%-----%------%
%-----%-----%------%

%-----%-----%------%%-----%-----%------%
\paragraph{Battery capital costs}
%-----%-----%------%%-----%-----%------%
Even though modern battery technology has been commercially available for several decades, their capital costs have dropped significantly over the last few years due to technological developments and scaling effects in the manufacturing process (see~\cite{Nykvist2015} for a review on the cost development of battery packs for electric vehicles). Budischak et al. in~\cite{budischak2013} estimate a price drop by roughly 70\% between 2008 and  2030 to values of 310~\euro/kW for power and 145~\euro/kWh for energy capacity, as used in the base scenarios.

\begin{figure}
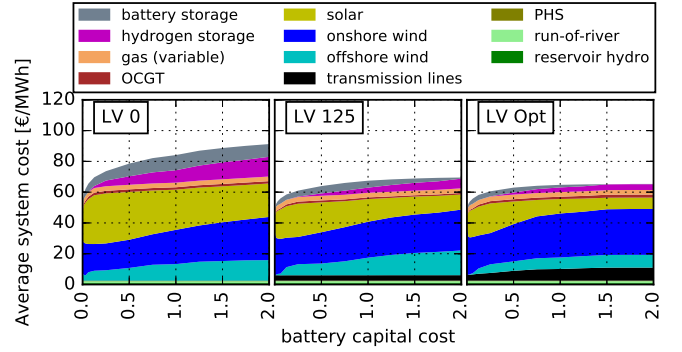
%[htbp]
\centering
\includegraphics[width=\linewidth]{{{diw2030_solar1_7_batterycapital/avgcosts_wHydroFOM_LVall_diw2030_solar1_7_batterycapital}}}
\caption{Composition of the average total system costs per unit of generated energy as a function of the fraction of the base battery capital cost  for the zero, compromise, and optimal (left to right panels) transmission grid scenarios.}
\label{fig:batterycapital_cost}
\end{figure}

Since in our model storage power and energy capacity are assumed to be coupled via the technology-dependent time scale $h_{s,max}$ for full charging / discharging, a battery cost reduction which uniformly affects both these cost components is assumed. Fig.~\ref{fig:batterycapital_cost} indicates that the modelling results presented here are quite robust against even large changes of battery costs down to 25\% relative to the base scenarios assumption. In that case, the total system costs decrease by only 10.6\% (7\%) to 61 \euro/MWh (60.6 \euro/MWh) in the compromise (optimal) grid scenario.
Without interconnecting transmission, storage has to provide a large amount of flexibility to ensure system stability.
Consequently, the total system costs are more sensitive to the storage costs and decrease by 14.3\% to 73.6 \euro/MWh for the same battery cost change.

%-----%-----%------%
%-----%-----%------%

\begin{figure}
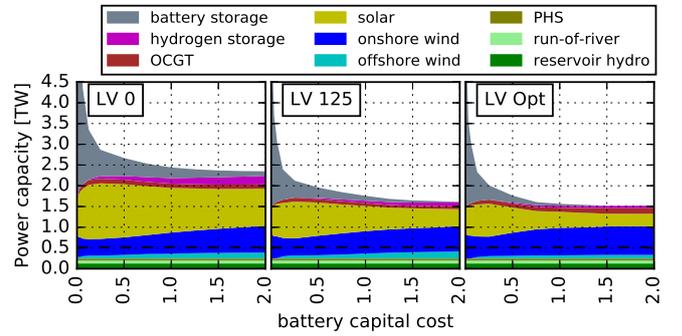
%[htbp]
\centering
\includegraphics[width=\linewidth]{{{diw2030_solar1_7_batterycapital/powercapa_wHydroFOM_LVall_diw2030_solar1_7_batterycapital}}}
\caption{Composition of the total installed power capacity of generators and storage units as a function of the fraction of the base battery capital cost assumption for the zero, compromise, and optimal (left to right panels) transmission grid scenarios. The peak of the total demand of 517 GW is marked as horizontal dashed black line. Recall that the battery energy capacities are given by the power capacities multiplied by the time scale $h_{\mathrm{battery},\mathrm{max}}=6 \mathrm{h}$.}
\label{fig:batterycapital_power}
\end{figure}

The installed battery power increases exponentially with decreasing cost from 0.12 TW to 0.41 TW
at 25\% of the costs (see Fig.~\ref{fig:batterycapital_power}).
However, the solar share increases only linearly at a low rate as more battery capacity becomes available, but stays below 35\% of the generated energy. The batteries allow more short-term smoothing and therefore more efficient usage of fluctuating solar generation. This also reduces the required wind capacities, and replaces to a larger extent the more capital intensive offshore wind power.
Additionally, the long-term H$_2$ storage can be completely removed due to a lower wind share and higher solar energy efficiency. The peak demand coverage provided in a few hours per year is then cost-efficiently replaced by additional OCGT power capacity.

%-----%-----%------%
%-----%-----%------%

For very small capital costs, the increase of the total battery power capacity as a function of the cost factor  becomes even stronger. This is due to the interaction with onshore wind power. In this limit, the collective battery energy capacity is large enough for long-term storage and allow to smooth out even wind fluctuations over at least several days. Onshore wind is the cheapest renewable generator type in the model in terms of investment costs per average producible energy and can therefore reduce the system costs by replacing solar and offshore wind installations if power balancing is no longer the limiting factor.

%-----%-----%------%%-----%-----%------%
\paragraph{Hydrogen storage capital costs}
%-----%-----%------%%-----%-----%------%
Hydrogen (H$_2$) storage that is charged by electrolysis of water and discharged via a fuel cell is not yet deployed at a large scale.
Its cost and efficiency parameters are characterised by expensive power capacity, but cheap energy capacity with lower round-trip efficiency when compared to battery storage. This makes it adequate for a long-term, large energy storage profile with relatively low usage frequency (see also the discussion in~\cite{budischak2013} which analyses the system benefit of storage options for a highly renewable electricity system). Since a commercial large scale implementation is not yet available, large deviations from the base scenario cost assumptions are plausible.

\begin{figure}
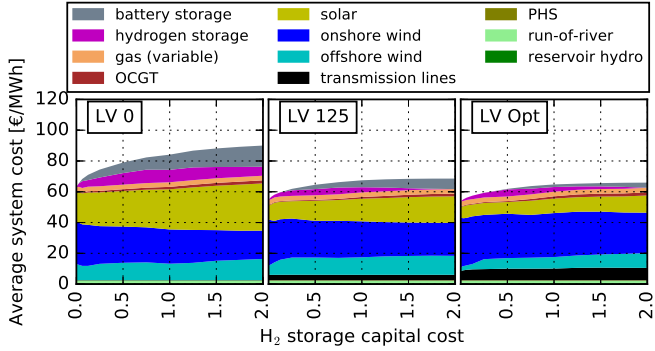
%[htbp]
\centering
\includegraphics[width=\linewidth]{{{diw2030_solar1_7_H2capital/avgcosts_wHydroFOM_LVall_diw2030_solar1_7_H2capital}}}
\caption{Composition of the average total system costs per unit of generated energy as a function of the fraction of the base H$_2$ storage capital cost assumption for the zero, compromise, and optimal (left to right panels) transmission grid scenarios.}
\label{fig:H2_cost}
\end{figure}

The modelling results are even more robust against changes of the H$_2$ storage costs than of the battery costs as shown in Fig.~\ref{fig:H2_cost}. Here again a uniform decrease in both power and energy capacity of the storage is assumed, which are coupled via the time scale $h_{H_{2},\mathrm{max}}=168\mathrm{h}$. The total system costs correspond closely to the ones in the previous section down to 25\% of the base cost assumptions, after which they fall slightly slower with decreasing H$_2$ storage costs.

%-----%-----%------%
%-----%-----%------%

\begin{figure}
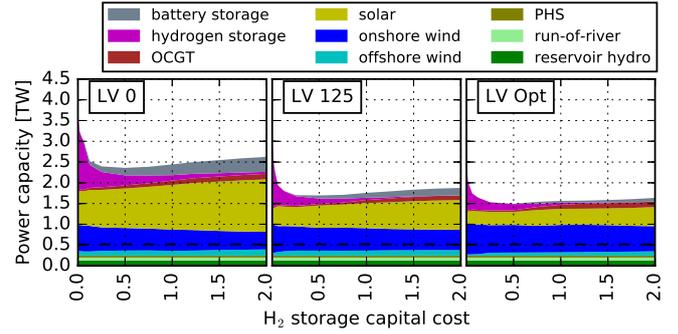
%[htbp]
\centering
\includegraphics[width=\linewidth]{{{diw2030_solar1_7_H2capital/powercapa_wHydroFOM_LVall_diw2030_solar1_7_H2capital}}}
\caption{Composition of the total installed power capacity of generators and storage units as a function of the fraction of the H$_2$ storage capital cost assumption relative to the base scenarios. The peak of the total demand of 517 GW is marked as horizontal dashed black line. Recall that the hydrogen storage energy capacities are given by the power capacities multiplied by the time scale $h_{\mathrm{H_{2}},\mathrm{max}}=168 \mathrm{h}$.}
\label{fig:H2_power}
\end{figure}

However, the composition of the system under decreasing H$_2$ storage costs remains almost identical even for a 75\% cost reduction that leads to a capacity increase by a factor of $4.3$ in the compromise transmission scenario (see Fig.~\ref{fig:H2_power}). The share of onshore wind increases only slightly as the H$_2$ storage cost is lowered. Onshore wind is already the dominant energy source in the system, but can still be made more efficient by additional long-term storage.
Simultaneously, some of the solar installations and eventually all battery capacities are removed.
Even though the round-trip efficiency of batteries is significantly higher, inexpensive and large H$_2$ storage capacities, in combination with increased wind installations, provide enough flexibility to the system to even compensate most short-term solar fluctuations.

%-----%-----%------%
%-----%-----%------%

The installed power capacity of the conventional gas turbines (OCGT) quickly decreases to almost zero already at 80\% H$_2$ costs, but never reaches the limit of zero capacity. The amount of generated energy from this source is determined by the CO$_2$ emission constraint.
The OCGT power capacity decreases because there is sufficiently large H$_2$ storage capacity which allows to cover all demand peaks mostly from wind energy that previously had to be curtailed but can now be stored in large quantities and over sufficiently long periods. In turn OCGT power does not have to be preserved for extreme hours, but is used by the model as a relatively cheap energy source that is operated almost continuously over the whole year and only shuts down during peak solar production. This maximises the capacity factor of OCGT and avoids most of its capital costs.

If H$_2$ storage is more expensive than in the base assumption, the previously discussed trends reverse. In this case, H$_2$ storage is at some point replaced by OCGT, battery, and solar capacity, and is no longer deployed if the costs are increased to more than 175\% for both the compromise and the optimum grid scenario.
Without transmission, storage is much more important for the security of supply for the system and 36\% of the H$_2$ storage capacity is built even if its cost doubles (see~\cite{schlachtberger_benefits_2017} for details).

%%%%%-----%%%%%-----%%%%%-----%%%%%------%%%%%
%%%%%-----%%%%%-----%%%%%-----%%%%%------%%%%%
\section{Results III: Influence of policy constraints}
\label{sec:constraints}
%%%%%-----%%%%%-----%%%%%-----%%%%%------%%%%%
%%%%%-----%%%%%-----%%%%%-----%%%%%------%%%%%
In~\cite{schlachtberger_benefits_2017} the role of different levels of constraints on transmission capacity expansion for cost-efficient layouts of a European electricity system is investigated. The limited geographic potential of different generation and storage technologies as well as a cap on CO$_{2}$ emissions entered as fixed constraints into the optimisation. However, for onshore wind a further restriction beyond geographical limits due to public acceptance issues is plausible. In a second investigation different CO$_2$ caps are implemented, representing less or even more ambitious emission targets than the 95\% reduction compared to 1990 levels assumed in the base scenario. The influence of these policy constraints on the system optimisation results are examined.

%%%%%-----%%%%%-----%%%%%-----%%%%%------%%%%%
\subsection{Onshore wind potentials}
%%%%%-----%%%%%-----%%%%%-----%%%%%------%%%%%
Despite a general positive public attitude towards renewable power generation (in~\cite{EUsurvey2017}, a survey of European Union citizens
for the European Commission in 2017, 89\% thought it was important for
their national government to set targets to increase renewable energy
use by 2030), local onshore wind energy projects appear to be facing increasing opposition throughout Europe (see~\cite{ellis_social_2016} for a discussion of the social acceptance of wind energy). Social constraints to the exploitation of onshore wind resources are represented by reducing their maximum installation capacity in each region to a fraction of the geographic potential.

\begin{figure}
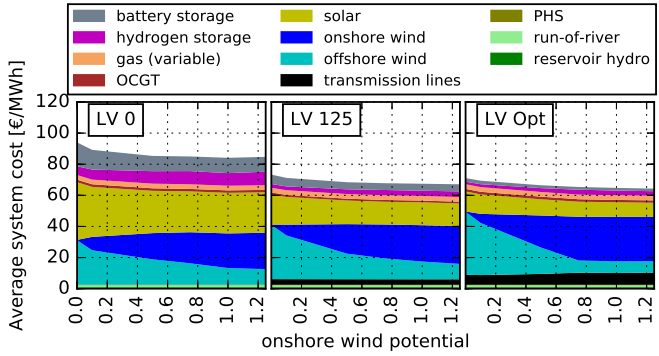
%[htbp]
%\centering
\includegraphics[width=\linewidth]{{{diw2030_solar1_7_owp/avgcosts_wHydroFOM_LVall_diw2030_solar1_7_owp}}}
\caption{Average total system costs per unit of generated energy in \euro/MWh as a function of the fraction of the onshore wind potential limit for the transmission line volumes of the zero, compromise, and optimal (left to right panels) grid.}
\label{fig:owp}
\end{figure}

Fig.~\ref{fig:owp} shows the optimisation results for scenarios with such a constrained onshore wind potential. It is found that in this case, onshore wind generation is almost completely replaced by offshore wind. This replacement is not linear with the reduction of onshore wind potential. Some of the onshore installations can be moved to other regions which were not fully using their potentials in the base scenario and have only slightly worse wind conditions.
Even though offshore wind turbines have assumed capital costs twice as high as those on land, their average capacity factor is usually much higher ($0.485$ vs. $0.233$ for a layout given by the installation potential $G_{n,s}^{max}$, see Sec.~\ref{sec:input_data}) and therefore the average energy generation costs are only slightly higher offshore than onshore.
This becomes apparent in the only small increase of total system cost as the amount of onshore wind installation potential is decreased down to zero.
The costs change by less than 2.6\% if the potential is reduced by half and only up to 8.8-12.2\% if no onshore wind is allowed, with the largest increase if transmission is also strongly limited.

%-----%-----%------%
%-----%-----%------%

Only for strongly reduced potentials do small changes in other parts of the system become visible.
Not all countries have access to the sea to build offshore turbines with sufficiently high capacity factors or installation potentials. If transmission capacities are restricted, they have to install slightly more solar PV and batteries, which leads to increasing costs. H$_2$ storage can be reduced by up to 41\% since the feed-in from offshore wind is smoother, but some long term storage is still cost-optimal and the overall cost effect is almost negligible.

\begin{figure}
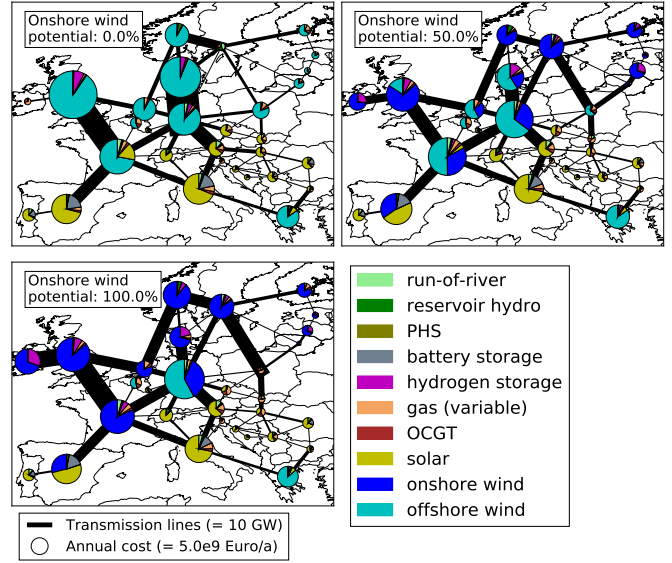
%[htbp]
\centering
\includegraphics[width=\linewidth]{{{diw2030_solar1_7_owp/network-euro-pie-costs_hydro-diw2030_solar1_7_owp-LVOpt-combined-0.0-0.5-1.0}}}
\caption{Map of average annual system costs per country in the optimal transmission scenario for three levels of onshore wind potential 0\% (top left), 50\% (top right), 100\% (bottom, base case) of the base assumption in each country.
The area of the circles is proportional to the total costs per country. The colors represent the shares of the different technologies. The modelled international transmission lines are shown as black lines with width proportional to their optimised net transfer capacity.
}
\label{fig:owp_cost_map}
\end{figure}

\begin{figure}
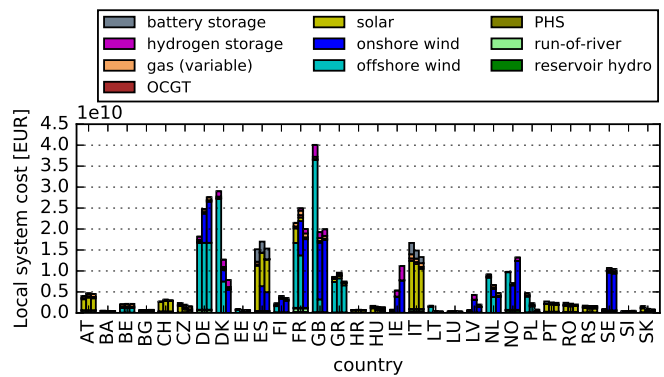
%[htbp]
\centering
\includegraphics[width=\linewidth]{{{diw2030_solar1_7_owp/costs_ct-diw2030_solar1_7_owp-LVOpt_0.0_0.5_1.0}}}
\caption{Same data as in Fig.~\ref{fig:owp_cost_map} but as a more quantitative bar plot with explicit values of the local system cost and its composition without representation of the transmission lines. The three bars for each country show the level of onshore wind potential 0\% (left bar), 50\% (middle bar), 100\% (right bar), respectively.}
\label{fig:owp_cost_cts}
\end{figure}

The same effects can be observed at country scale as shown in Figs.~\ref{fig:owp_cost_map} and \ref{fig:owp_cost_cts} for the case with optimal transmission. Onshore wind capacity is replaced by offshore wind in countries bordering the North and Baltic Sea, which is significantly extended in Denmark, the Netherlands, and eventually also in Great Britain. Germany already builds the maximum installable offshore wind capacity in the base scenario and therefore does not replace its onshore wind power. Ireland also does not replace its onshore wind, but simply reduces it. The very large transmission capacities that connect Great Britain and Denmark to their southern neighbours suggest that the energy can now be produced efficiently and without a large grid connection to Ireland in more central nodes of the network.
Most countries in the southern half of Europe retain or slightly increase their solar capacity. If no onshore wind is allowed, France installs 45 GW of solar power but no local battery or H$_2$ storage, while most other countries expand their battery storage together with the solar capacity. This suggests that the solar PV fluctuations in France can be balanced by the more continuous offshore wind generation locally or through imports from the strong grid connection with Great Britain.

%-----%-----%------%
%-----%-----%------%

%-----%-----%------%
%-----%-----%------%

The results presented in this study build on a coarse-grained model with each country aggregated to a single node. Onshore potentials and in particular offshore potentials are heterogeneously distributed inside the countries. Corresponding intra-country transmission needs are ignored due to the spatial scale of the model. Nevertheless, since transmission costs represent only a minor component of the total system costs, our results are expected to be stable on a finer scale as long as intra-country transmission can be expanded freely. However, if intra-country capacities are restricted, for example due to public acceptance issues, it may not be possible to integrate all the energy from wind power plants. The reader is referred to the results in~\cite{horsch_role_2017} and a forthcoming study by the same authors.

%%%%%-----%%%%%-----%%%%%-----%%%%%------%%%%%
\subsection{Carbon dioxide emission constraint}
%%%%%-----%%%%%-----%%%%%-----%%%%%------%%%%%
Setting a limit to the total European CO$_2$ emissions is motivated by the policy goal to keep global temperatures from increasing beyond a certain level. In 2011 the European Council reconfirmed the EU objective to reduce overall greenhouse gas emissions by 80-95\% compared to 1990 values, including a corresponding emission decrease in the power sector of 93-99\%~\cite{eu2050}. In the electricity system model applied in the present study, for the base scenario this target is represented as a CO$_2$ emission limit $CAP_{CO2} = 5\%$ in units of the emission level in the year 1990, i.e., 77.5 Mt-CO$_2$-equivalent per year for the electricity sector (this value is derived from the data given in~\cite{ EEACO22017}, see~\cite{schlachtberger_benefits_2017} for details).

Going beyond this limit, CO$_2$ emissions can be brought down to zero by replacing the remaining fossil-fuel based conventional generation. The only conventional generator and thus source of CO$_2$ emissions in the model are open-cycle gas turbines (OCGT). This type of generation capacity is highly flexible and has relatively low investment costs for power capacity, but high variable costs for energy generation. This  makes this technology option well suited to cover peak residual demand in a few hours per year (see Tab.~\ref{tab:results_costs} for the respective cost assumptions).

\begin{figure}
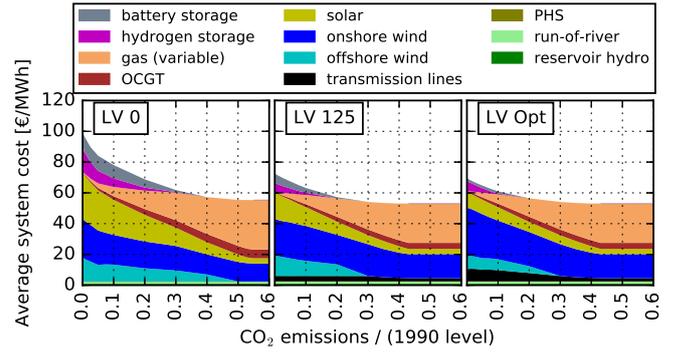
%[htbp]
\centering
\includegraphics[width=\linewidth]{{{diw2030_solar1_7_co2/avgcosts_wHydroFOM_LVall_diw2030_solar1_7_co2_extended_margsep}}}
\caption{Average total system costs per unit of generated energy as a function of the CO$_2$ emission limit defined as fraction of the emission level of the year 1990, i.e. 1.55~Gtonne-CO$_2$, for the zero, compromise, and optimal transmission grid (left to right panels). The base scenarios assume a CO$_2$ emission level of 5\%. For a CO$_2$ emission level above 30\%, the cost-optimal line volume is below 125 TWkm and therefore the compromise and optimal transmission cases are identical.}
\label{fig:co2ext}
\end{figure}

Fig.~\ref{fig:co2ext} shows that the total system costs are an almost linear function of the CO$_2$ emission limit close to $CAP_{CO2}=5\%$ with an approximate rate of $-0.94$ and $-0.86$ \euro/MWh per percentage point of $CAP_{CO2}$ for compromise and optimal transmission volume, respectively. With the compromise (optimal) grid, a system with zero emissions has average system costs of 72 (69) \euro/MWh, a 6.6\% (6.4\%) increase from the base scenario.

%-----%-----%------%
%-----%-----%------%

In the other limit, for an CO$_2$ emission level above 30\%, the transmission system expansion constraint for the compromise grid scenario is no longer binding, i.e. the line volume in the cost-optimal system set-up is below 125 TWkm, leading to identical solutions for both transmission scenarios. The CO$_2$ constraint in this case is binding up to an emission level of approximately 43\%. Since the only emissions in the model result from burning the fuel in the gas turbines, in this range the corresponding component of the system costs in Fig.~\ref{fig:co2ext} can directly be calculated, using the allowed total CO$_2$ emissions, the CO$_2$ emission intensity and efficiency of the gas turbines, and the fuel costs. The gas turbine fuel component of the average system costs is then approximately $CAP_{CO2}\times 55.39$ \euro/MWh, with $CAP_{CO2}$ denoting the CO$_2$ emission constraint relative to 1990 levels. For $CAP_{CO2}=43\%$ or 0.67 Gtonne-CO$_2$/year this corresponds to 23.82 \euro/MWh (see Fig.~\ref{fig:co2ext}).

Once the emission constraint becomes binding, the system optimisation is forced to reduce the amount of energy generated from gas turbines. Fig.~\ref{fig:co2ext} shows that down to about 20\% the missing energy is then delivered by increasing renewable generation capacity, mostly from solar and offshore wind, while the generation capacity of the gas turbines remains constant. At some point (around 20\%/15\% for the compromise/optimal grid scenario), the restriction in the usage of the generation from gas turbines starts to affect situations of peaks in the residual load, for which a corresponding increase in renewable generation is no longer a cost-efficient solution. Instead, the system optimisation has to provide the necessary flexibility by a combination of battery and hydrogen storage, which replaces the respective dispatchable power generation also in other situations (see the  decrease of gas capacity system costs in Fig.~\ref{fig:co2ext}). Accordingly, the sum of the installed power capacity of OCGT, battery, and H$_2$ storage is almost constant at $49 \pm 3 \%$ of the peak demand for the compromise grid. Similarly, for the optimal grid the total power of OCGT, battery, H$_2$ storage, and offshore wind is $52 \pm 2 \%$ of peak demand, relatively independent of the emission level. In the latter case, the optimal transmission capacity allows a system-wide usage of the comparatively steady power generation from offshore wind. Nevertheless, the replacement of gas turbines by storage options for stricter emission constraints leads to a stronger increase in system costs, caused by the higher capital costs and reduced efficiencies of these technologies (see Tab.~\ref{tab:results_costs} for the respective cost assumptions and efficiencies). Furthermore, the losses associated with using storage technologies represent an additional factor for increasing renewable generation capacities and costs.

Fig.~\ref{fig:co2ext} shows that the effects of a varying CO$_2$ emission limit are even more pronounced in the zero transmission scenario. Without transmission the emission constraint becomes binding at $CAP_{CO2}\approx 53\%$. Power generation from gas is then replaced by generation from larger renewable capacities. At $CAP_{CO2}<30\%$, storage units start to replace gas turbines. Transmission not only provides a smoothing effect with respect to the fluctuating renewable generation, but also allows a more efficient system-wide usage of the flexibility provided by storage technologies. Correspondingly, for the zero transmission scenario the total power of OCGT, battery, and H$_2$ storage increases with a decreasing emission limit. In particular, in the limit $CAP_{CO2}<5\%$ both the costs for renewable generators and the storage units grow exponentially.

Open cycle gas turbines are assumed to be the only conventional technology in the system due to their very high flexibility. This assumption starts to break down once conventional generation is no longer needed only for peak load coverage. Today, other conventional technologies, e.g. modern coal power plants, can be more economically efficient if less flexibility and more continuous bulk generation is required. This would lead to a slight reduction of total costs in the extreme case of large CO$_2$ emission limits. A more thorough analysis of this effect would require detailed modelling of ramping constraints and costs and lies beyond the focus of this work (see~\cite{schlachtberger_backup_2016} for a discussion of ramping constraints in a model similar to the one used in~\cite{schlachtberger_benefits_2017}).

\begin{figure}
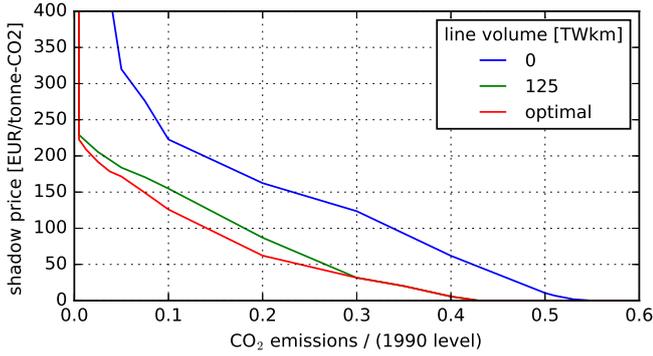
%[htbp]
\centering
\includegraphics[width=\linewidth]{{{diw2030_solar1_7_co2/shadow_price_co2_diw2030_solar1_7_co2}}}
\caption{Shadow price of the global CO$_2$ emission constraint as a function of the CO$_2$ emission limit $CAP_{CO2}$ for the zero, compromise, and optimal (blue, green, red lines) transmission scenario. For exactly zero emissions, the shadow price diverges to around 20,000 \euro/tonne-CO$_2$, two orders of magnitude larger than the plotted range. The green and red lines are partly overlapped.}
\label{fig:shadowprice_co2}
\end{figure}

%-----%-----%------%%-----%-----%------%
\paragraph{Carbon dioxide emission shadow price}
%-----%-----%------%%-----%-----%------%
It should be emphasised that the results above assume a direct CO$_2$ emission price of 0 \euro/tonne-CO$_2$. Higher prices would lead to smaller economically optimal OCGT shares and emission levels. Fig.~\ref{fig:shadowprice_co2} plots the shadow price $\mu_{CO2}$ of the global CO$_2$ emission constraint in equation \eqref{eq:co2cap} against $CAP_{CO2}$. Here, $\mu_{CO2}$ can be interpreted as the CO$_2$ price that would be required for the market to achieve a certain emission level under the given assumptions (for a more general discussion of shadow prices in the context of efficient electricity markets see~\cite{biggar2014economics}).

%-----%-----%------%
%-----%-----%------%

A restriction to the $CAP_{CO2} = 5\%$ level of the base scenario is economically optimal if the emission price is set to roughly 180 \euro/tonne-CO$_2$ for both compromise and optimal grid.
In these transmission scenarios, the effect of the emission price is similar and the change of $\mu_{CO2}$ with $CAP_{CO2}$ is close to linear down to very small emission limits. Only if no emissions are allowed, the CO$_2$ shadow price jumps to around 20000 \euro/tonne-CO$_2$, two orders of magnitude larger than the previous values. This suggests that there is a small number of hours per year when the flexibility provided by the power capacity of OCGT is very valuable for system stability. The system has to install significant additional capacities only to cover the demand in these few hours. In practice, load shedding would be a more viable option for these rare events, but is not included in the model.

The flexibility from the dispatchable generators is much more valuable if fluctuations cannot be smoothed by transmission. In this case, $\mu_{CO2}$ grows at a similar rate with decreasing CO$_2$ limit as with transmission but is higher by 60 to 100\,\euro/tonne-CO$_2$ down to $CAP_{CO2} > 10\%$. For stricter emission limits, $\mu_{CO2}$ increases significantly faster at an exponential rate. $CAP_{CO2} = 5\%$ can be obtained by an emission price of 319\,\euro/tonne-CO$_2$, while $CAP_{CO2} = 0.5\%$ requires $\mu_{CO2}=1280$\,\euro/tonne-CO$_2$. This underlines how expensive an ambitious CO$_2$ reduction is when grid capacity is severely limited.

%-----%-----%------%%-----%-----%------%
\paragraph{CO$_2$ emission constraint without storage}
%-----%-----%------%%-----%-----%------%
In the previous section it was shown that installing battery and H$_2$ storage is cost-efficient only if the CO$_2$ emission limit is very low. In the following the consequences of removing storage altogether are examined, both to understand better the economic necessity of storage and to analyse the case where unforeseen feasibility problems hinder the large-scale deployment of storage.

\begin{figure}
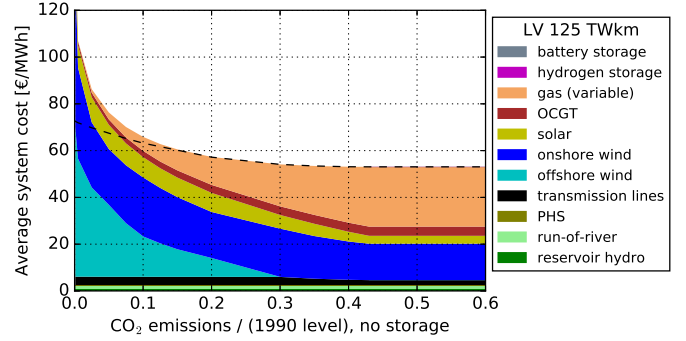
%[htbp]
\centering
\includegraphics[width=\linewidth]{{{diw2030_solar1_7_nostor_co2/avgcosts_wHydroFOM_LV0.25_diw2030_solar1_7_nostor_co2_extended_margsep}}}
\caption{Same as Fig.~\ref{fig:co2ext} for the compromise grid but without battery and H$_2$ storage included in the model. The total system costs for the case including both storage technologies is marked with a black dashed line. For emission levels above 20\%, the results are identical to Fig.~\ref{fig:co2ext}. The optimisation did not converge for zero emissions. The smallest computed level of 0.1\% has a total cost of 149.6 \euro/MWh (above plotted range).
}
\label{fig:nostor_co2}
\end{figure}

Consult Fig.~\ref{fig:nostor_co2} for the cost development as the CO$_2$ constraint is restricted for the case of moderate transmission expansion (compromise grid scenario). As outlined in the last section, for CO$_2$ emission reductions to levels above 20\%, the results do not differ from the corresponding optimisation including storage. Below 20\% the costs rise much faster, being 13.1\% higher at 76.3 \euro/MWh for 5\% CO$_2$, 48.6\% higher at 106.9 \euro/MWh for 0.5\% CO$_2$ and infeasible for 0\% CO$_2$. There is also a significant increase in offshore wind in the range of low CO$_2$ emissions, as the system optimisation tries to exploit the lower fluctuations of offshore wind to cover the peak residual load.

%-----%-----%------%
%-----%-----%------%

The reason for these strongly increasing costs is the difficulty of bridging times with low wind and sun when there is neither flexibility provided by storage nor gas power generation. Countries must either considerably expand their wind and solar capacities to compensate for the low power availability, or import from other regions where better renewable resource quality provides more generation availability during these times (but even this is restricted by the moderate transmission capacity). The model increasingly introduces offshore wind because of its more regular production characteristics, but as CO$_2$ emissions are reduced, this is not sufficient to cover the demand at times with strongly adverse weather conditions for wind and solar power generation.

%%%%%-----%%%%%-----%%%%%-----%%%%%------%%%%%
%%%%%-----%%%%%-----%%%%%-----%%%%%------%%%%%
\section{Discussion: Limitations of the study}
\label{sec:discussion}
%%%%%-----%%%%%-----%%%%%-----%%%%%------%%%%%
%%%%%-----%%%%%-----%%%%%-----%%%%%------%%%%%
This contribution studies the sensitivity of cost optimal scenarios to various influences for the model presented in~\cite{schlachtberger_benefits_2017}. The investigations are explicitly not extended to alternative models, which might apply different methodological approaches, incorporate other sectors (like heating and transport) and other technologies (for instance nuclear generation or carbon capture), or consider a finer temporal or spatial scale. Such alternative models will show different sensitivities, depending on the specific modelling set-up  (see for instance~\cite{bjelic2015} for a discussion of  the sensitivity to energy and emission market prices for an optimisation model for sustainable energy systems). For models applying heuristic solutions, the behaviour of the cost function around the optimum -- which can be assessed using a multi-parameter sensitivity analysis -- will also have an influence on the necessity to adopt more elaborated but computationally expensive optimisation methods (see~\cite{bjelic2016} for a comparison of heuristic with optimisation methods in the context of an energy planning tool for sustainable national energy systems). Nevertheless, the restriction to a single model, for which there is complete knowledge of the underlying data and control over all modelling details, allows to assess comprehensively the model-inherent sensitivities to input data and optimisation constraints. In contrast, an inter-model comparison would have to compare scenarios derived from different models for common objectives, without necessarily taking into account the details of the modelling process. For the limitations of the underlying electricity system model the reader is referred to the discussion in~\cite{schlachtberger_benefits_2017}.

The strict restriction to one model is applied in Sec.~\ref{sec:input_data}, where the influence of the sampling in the input load and weather (i.e.  renewable generation) time series is studied. Here different samples from the same source are chosen and accordingly alternative sources or modelling frameworks are not considered. The load data is at country-level as the spatial scale of the present model is given from the transmission system operators (ENTSO-E), so a dependence on the regionalisation of the load data as would be necessary in spatially more detailed system models can be excluded (see~\cite{horsch_role_2017} for a discussion of the role of spatial scale in the optimisation of energy system models). Nevertheless, the load data could change both in its profile and volume under an increasing electrification and coupling to other energy sectors, for instance due to a wider use of heat pumps or electric vehicles. Sector coupling also enables new sources of flexibility, such as demand-side management of charging battery electric vehicles, or the use of cheap energy storage in the heating and the gas sectors, which may change the sensitivity of the system to the input parameters, particularly with regard to storage technologies. Such `smart energy systems' were presented and analysed in \cite{Lund2010}, while the interaction with cross-border transmission reinforcement was studied for a variety of flexibility scenarios by some of the authors in \cite{Brown2018}. Since low-fossil sector-coupling scenarios typically require significant electrification of other energy sectors, and wind and solar power still dominate electricity provision, many of the present conclusions with respect to wind and solar input data are expected to hold.

For the renewable generation potentials, both an influence from the underlying weather data source and from the modelling approach used for the conversion from weather to generation time series is expected (see~\cite{staffell_using_2016} and~\cite{Andresen2015} for a discussion of reanalysis models for wind power output data). Also climate change will have an influence on the structure of future weather conditions, a topic which recently has been discussed for instance in~\cite{Hdidouan2017,Wohland2017,Karnauskas2018}. Given the importance of the spatio-temporal renewable generation patterns for a highly renewable electricity system, these topics need to be addressed extensively in future research, but are beyond the scope of the present study.

The influence of cost assumptions is discussed in Sec.~\ref{sec:cost}, where the variation of one cost component is considered while the remaining parameters are fixed. For the sensitivity to multi-parameter variations a linear behaviour for small changes is expected, i.e. an overlay of the individual system effects. For larger variations, non-linear effects will lead to new cost optimal system configurations. Although an extensive exploration of the whole parameter space is not feasible in the context of a single journal paper, it is assumed that the essential interdependencies are already covered in the single-parameter variation approach.
%%%%%-----%%%%%-----%%%%%-----%%%%%------%%%%%
%%%%%-----%%%%%-----%%%%%-----%%%%%------%%%%%
\section{Summary and Conclusions}
\label{sec:conclusion}
%%%%%-----%%%%%-----%%%%%-----%%%%%------%%%%%

%%%%%-----%%%%%-----%%%%%-----%%%%%------%%%%%
Models of the electricity system give important insights into how to cost-efficiently combine different technology options in the framework given by the physical, environmental, or societal constraints of the system. Even if the methodological approach and the scope of a model is fixed, the simulation results will depend on the assumptions concerning the input data, input parameters, and constraints inside the model. Using the techno\hyp{}economic optimisation model for the European electricity system presented in~\cite{schlachtberger_benefits_2017}, the influence of the data sampling in the load and renewable generation time series, and of different cost assumptions for the capital costs of generation and storage technologies is studied. Beyond this analysis, it is shown how different policy constraints, in particular the cap on the CO$_2$ emission level, affect the structure of the cost optimal scenarios.

It is observed that trends in different samples of the load and renewable generation time series are reflected in the details of the simulation results, but only weakly affect the resulting total system costs. The robustness of the overall results to input weather data from different years confirms earlier results from the literature \cite{Lund2003}, however specific technologies may be affected by inter-annual variability, so the most robust results are found by taking weather data from several years (see for instance \cite{Collins2018}). A shift from an hourly resolution to a 3 hour sampling in the time series can increase the share of solar power generation due to the corresponding smoothing of fluctuations on this time scale, but again the changes are slight.

For moderate changes in the solar capital costs a linear relationship to total system costs is observed, with onshore wind power being replaced by additional solar and battery capacity for decreasing solar costs. A comparatively stronger sensitivity to onshore wind capital costs is shown, which can be explained by the systemic importance of wind in meeting the critical peak winter demand in Europe. Given a scenario with moderate transmission expansion, a reduction of onshore wind capital costs by 25\% leads to a decrease of system costs by 10.4\%. In contrast, a reduction of battery or hydrogen storage capital costs only has a very weak effect on the modelling results. All in all for moderate changes in the cost assumptions it is observed, that the total system costs tend to be flat in the optimisation space, which indicates a certain degree of freedom to consider additional factors like public acceptance in the choice of cost optimal system layouts. The consideration of constraints in the exploitation of onshore wind potentials leads to similar results. In this case, onshore wind capacity is mostly replaced by offshore wind capacity, with only a small increase in total system costs.

The base scenarios studied in~\cite{schlachtberger_benefits_2017} assume a CO$_2$ emission limit of 5\% in terms of 1990 levels as a constraint for the system optimisation. Since open cycle gas turbines are the only source of CO$_2$ emissions considered in the model, this constraint directly translates into the amount of flexible power generation from natural gas. The simulation results shows that a 5\% cap on emissions corresponds to a CO$_2$ shadow price of 180 \euro/(tonne-CO$_2$) for both a scenario with optimal and with moderate transmission expansion. This shadow price indicates the carbon dioxide price necessary to obtain the corresponding reduction in emissions in an unconstrained market. For a scenario without transmission between the European countries the CO$_2$ shadow price rises to 319 \euro/(tonne-CO$_2$), underlining the benefit of transmission for a low-emission electricity system~\cite{schlachtberger_benefits_2017}. It is observed that for both grid expansion scenarios the emission limit becomes binding for a 43\% reduction compared to 1990s level. Stricter constraints necessarily lead to a decrease in the share of power generated from natural gas, which is replaced by renewable generation from expanded wind and solar generation capacities at slightly increasing system costs. For even lower emissions, storage options replace gas turbines and provide the flexibility to meet peak demand situations in cost optimal scenarios. If such storage options are excluded from the system, total costs rise significantly, and the extreme case of a zero-emission scenario becomes infeasible.

This study addresses the sensitivity to changes in the input parameters and to policy constraints for  cost optimal scenarios of a low emission electricity system, highlighting the stability of total system costs and the decisive role of the CO$_2$ emission constraint. Such an analysis provides an understanding of the mechanisms underlying the cost-efficient combinations of different technologies, thus providing insights beyond a mere listing of scenarios derived from running extensive simulations. Given the uncertainties and complexities of the energy system, such a perspective should be added also to more detailed system models, assessing the role of the spatio-temporal scales, technology options, or sectors considered in their implementation. It would also be desirable to develop analytic results, either from simplified models that maintain the system's sensitivities or from more detailed mathematical analysis of the solution space, thereby allowing a deeper understanding of the system interactions. Only such a broad methodological approach can provide the robust and comprehensible policy guidelines necessary to facilitate a cost-efficient transition to a low-emission sustainable energy system.

\section*{Acknowledgements}
The project underlying this report was supported by the German Federal Ministry of Education and Research under grant no.~03SF0472C. Mirko Sch\"{a}fer is partially funded by the Carlsberg Foundation Distinguished Postdoctoral Fellowship. David Schlachtberger, Tom Brown and Martin Greiner are partially funded by the RE-INVEST project (Renewable Energy Investment Strategies -- A two-dimensional interconnectivity approach), which is supported by Innovation Fund Denmark (6154-00033B). Tom Brown also acknowledges funding from the Helmholtz Association under grant no.~VH-NG-1352. The responsibility for the contents lies with the authors.

\bibliographystyle{elsarticle-num}
\biboptions{sort}
\bibliography{literatur}

\begin{thebibliography}{10}
\expandafter\ifx\csname url\endcsname\relax
  \def\url#1{\texttt{#1}}\fi
\expandafter\ifx\csname urlprefix\endcsname\relax\def\urlprefix{URL }\fi
\expandafter\ifx\csname href\endcsname\relax
  \def\href#1#2{#2} \def\path#1{#1}\fi

\bibitem{eu2050}
{{European Commission}},
  \href{http://eur-lex.europa.eu/legal-content/EN/TXT/PDF/?uri=CELEX:52011DC0112&from=EN}{{A
  roadmap for moving to a competitive low carbon economy in 2050}}, Tech. rep.,
  {{European Commission}} (Mar. 2011).
\newline\urlprefix\url{http://eur-lex.europa.eu/legal-content/EN/TXT/PDF/?uri=CELEX:52011DC0112&from=EN}

\bibitem{IRENA2018a}
IRENA,
  \href{http://irena.org/-/media/Files/IRENA/Agency/Publication/2018/Jan/IRENA_2017_Power_Costs_2018.pdf}{{Renewable
  Power Generation Costs in 2017}}, {International Renewable Energy Agency},
  {Abu Dhabi}, 2018.
\newline\urlprefix\url{http://irena.org/-/media/Files/IRENA/Agency/Publication/2018/Jan/IRENA_2017_Power_Costs_2018.pdf}

\bibitem{IRENA2018b}
IRENA, {European Commission},
  \href{http://www.irena.org/-/media/Files/IRENA/Agency/Publication/2018/Feb/IRENA_REmap_EU_2018.pdf}{{Renewable
  Energy Prospects for the {Europe}an Union}}, {International Renewable Energy
  Agency}, {Abu Dhabi}, 2018.
\newline\urlprefix\url{http://www.irena.org/-/media/Files/IRENA/Agency/Publication/2018/Feb/IRENA_REmap_EU_2018.pdf}

\bibitem{Rasmussen2012}
M.~G. Rasmussen, G.~B. Andresen, M.~Greiner, Storage and balancing synergies in
  a fully or highly renewable pan-{Europe}an power system, Energy Policy 51
  (2012) 642--651.
\newblock \href {http://dx.doi.org/10.1016/j.enpol.2012.09.009}
  {\path{doi:10.1016/j.enpol.2012.09.009}}.

\bibitem{Connolly2012}
D.~Connolly, H.~Lund, B.~Mathiesen, E.~Pican, M.~Leahy, The technical and
  economic implications of integrating fluctuating renewable energy using
  energy storage, Renewable Energy 43~(0) (2012) 47--60.
\newblock \href {http://dx.doi.org/10.1016/j.renene.2011.11.003}
  {\path{doi:10.1016/j.renene.2011.11.003}}.

\bibitem{budischak2013}
C.~Budischak, D.~Sewell, H.~Thomson, L.~Mach, D.~E. Veron, W.~Kempton,
  Cost-minimized combinations of wind power, solar power and electrochemical
  storage, powering the grid up to 99.9\% of the time, Journal of Power Sources
  225 (2013) 60--74.
\newblock \href {http://dx.doi.org/10.1016/j.jpowsour.2012.09.054}
  {\path{doi:10.1016/j.jpowsour.2012.09.054}}.

\bibitem{Cebulla2017}
F.~Cebulla, T.~Naegler, M.~Pohl, {Electrical energy storage in highly renewable
  {Europe}an energy systems: Capacity requirements, spatial distribution, and
  storage dispatch}, Journal of Energy Storage 14 (2017) 211--223.
\newblock \href {http://dx.doi.org/10.1016/j.est.2017.10.004}
  {\path{doi:10.1016/j.est.2017.10.004}}.

\bibitem{Czisch}
G.~Czisch,
  \href{http://nbn-resolving.de/urn:nbn:de:hebis:34-200604119596}{Szenarien zur
  zuk\"unftigen {S}tromversorgung}, Ph.D. thesis, Universit\"at Kassel (2005).
\newline\urlprefix\url{http://nbn-resolving.de/urn:nbn:de:hebis:34-200604119596}

\bibitem{Schaber2}
K.~Schaber, F.~Steinke, P.~M{\"{u}}hlich, T.~Hamacher, Parametric study of
  variable renewable energy integration in {E}urope: Advantages and costs of
  transmission grid extensions, Energy Policy 42 (2012) 498--508.
\newblock \href {http://dx.doi.org/10.1016/j.enpol.2011.12.016}
  {\path{doi:10.1016/j.enpol.2011.12.016}}.

\bibitem{Gils2017}
H.~C. Gils, Y.~Scholz, T.~Pregger, D.~{Luca de Tena}, D.~Heide, {Integrated
  modelling of variable renewable energy-based power supply in Europe}, Energy
  123 (2017) 173 -- 188.
\newblock \href {http://dx.doi.org/10.1016/j.energy.2017.01.115}
  {\path{doi:10.1016/j.energy.2017.01.115}}.

\bibitem{schlachtberger_benefits_2017}
D.~Schlachtberger, T.~Brown, S.~Schramm, M.~Greiner, The benefits of
  cooperation in a highly renewable {European} electricity network, Energy 134
  (2017) 469--481.
\newblock \href {http://dx.doi.org/10.1016/j.energy.2017.06.004}
  {\path{doi:10.1016/j.energy.2017.06.004}}.

\bibitem{pfenninger_energy_2014}
S.~Pfenninger, A.~Hawkes, J.~Keirstead, Energy systems modeling for
  twenty-first century energy challenges, Renewable and Sustainable Energy
  Reviews 33 (2014) 74--86.
\newblock \href {http://dx.doi.org/10.1016/j.rser.2014.02.003}
  {\path{doi:10.1016/j.rser.2014.02.003}}.

\bibitem{pfenninger_opening_2018}
S.~Pfenninger, L.~Hirth, I.~Schlecht, E.~Schmid, F.~Wiese, T.~Brown, C.~Davis,
  M.~Gidden, H.~Heinrichs, C.~Heuberger, S.~Hilpert, U.~Krien, C.~Matke,
  A.~Nebel, R.~Morrison, B.~M\"{u}ller, G.~Ple{\ss}mann, M.~Reeg, J.~C.
  Richstein, A.~Shivakumar, I.~Staffell, T.~Tr\"{o}ndle, C.~Wingenbach, Opening
  the black box of energy modelling: {Strategies} and lessons learned, Energy
  Strategy Reviews 19 (2018) 63--71.
\newblock \href {http://dx.doi.org/10.1016/j.esr.2017.12.002}
  {\path{doi:10.1016/j.esr.2017.12.002}}.

\bibitem{Pfenninger2017b}
S.~Pfenninger, J.~DeCarolis, L.~Hirth, S.~Quoilin, I.~Staffell, {The importance
  of open data and software: Is energy research lagging behind?}, Energy Policy
  101~(July 2016) (2017) 211--215.
\newblock \href {http://dx.doi.org/10.1016/j.enpol.2016.11.046}
  {\path{doi:10.1016/j.enpol.2016.11.046}}.

\bibitem{Schaber}
K.~Schaber, F.~Steinke, T.~Hamacher, Transmission grid extensions for the
  integration of variable renewable energies in {E}urope: Who benefits where?,
  Energy Policy 43 (2012) 123--135.
\newblock \href {http://dx.doi.org/10.1016/j.enpol.2011.12.040}
  {\path{doi:10.1016/j.enpol.2011.12.040}}.

\bibitem{PyPSA}
T.~Brown, J.~H\"orsch, D.~Schlachtberger, {PyPSA: Python for Power System
  Analysis}, Journal of Open Research Software 6~(1) (2018) 4.
\newblock \href {http://arxiv.org/abs/1707.09913} {\path{arXiv:1707.09913}},
  \href {http://dx.doi.org/10.5334/jors.188} {\path{doi:10.5334/jors.188}}.

\bibitem{schroeder2013}
A.~Schr\"{o}der, F.~Kunz, J.~Meiss, R.~Mendelevitch, C.~von Hirschhausen,
  \href{http://www.diw.de/documents/publikationen/73/diw_01.c.424566.de/diw_datadoc_2013-068.pdf}{Current
  and prospective costs of electricity generation until 2050}, Data
  Documentation, DIW~68, Deutsches Institut f\"{u}r Wirtschaftsforschung (DIW),
  Berlin (2013).
\newline\urlprefix\url{http://www.diw.de/documents/publikationen/73/diw_01.c.424566.de/diw_datadoc_2013-068.pdf}

\bibitem{Hagspiel}
S.~Hagspiel, C.~J{\"a}gemann, D.~Lindenberger, T.~Brown, S.~Cherevatskiy,
  E.~Tr{\"o}ster, Cost-optimal power system extension under flow-based market
  coupling, Energy 66 (2014) 654--666.
\newblock \href {http://dx.doi.org/10.1016/j.energy.2014.01.025}
  {\path{doi:10.1016/j.energy.2014.01.025}}.

\bibitem{ENTSOEinstalledcapas}
{European Transmission System Operators},
  \href{https://transparency.entsoe.eu/generation/r2/installedGenerationCapacityAggregation/show}{{Installed
  Capacity per Production Type in 2015}} (2016).
\newline\urlprefix\url{https://transparency.entsoe.eu/generation/r2/installedGenerationCapacityAggregation/show}

\bibitem{kies2016}
A.~Kies, K.~Chattopadhyay, L.~von Bremen, E.~Lorenz, D.~Heinemann, {RESTORE
  2050~{W}ork Package Report D12: Simulation of renewable feed-in for power
  system studies.}, Tech. rep., Carl von Ossietzky Universit\"{a}t Oldenburg,
  Germany (2016).

\bibitem{entsoe_load}
{European Transmission System Operators},
  \href{https://www.entsoe.eu/data/data-portal/consumption/}{{Country-specific
  hourly load data}} (2011).
\newline\urlprefix\url{https://www.entsoe.eu/data/data-portal/consumption/}

\bibitem{Heide2010}
D.~Heide, L.~von Bremen, M.~Greiner, C.~Hoffmann, M.~Speckmann, S.~Bofinger,
  Seasonal optimal mix of wind and solar power in a future, highly renewable
  {Europe}, Renewable {En}ergy 35~(11) (2010) 2483--2489.
\newblock \href {http://dx.doi.org/10.1016/j.renene.2010.03.012}
  {\path{doi:10.1016/j.renene.2010.03.012}}.

\bibitem{Andresen2015}
G.~B. Andresen, A.~A. S{\o}ndergaard, M.~Greiner, Validation of {D}anish wind
  time series from a new global renewable energy atlas for energy system
  analysis, Energy 93 (2015) 1074--1088.
\newblock \href {http://dx.doi.org/10.1016/j.energy.2015.09.071}
  {\path{doi:10.1016/j.energy.2015.09.071}}.

\bibitem{corine2006}
{European Environment Agency},
  \href{https://www.eea.europa.eu/data-and-maps/data/clc-2006-vector-4}{Corine
  land cover 2006} (2014).
\newline\urlprefix\url{https://www.eea.europa.eu/data-and-maps/data/clc-2006-vector-4}

\bibitem{natura2000}
{European Environment Agency},
  \href{http://www.eea.europa.eu/data-and-maps/data/natura-7}{Natura 2000 data
  - the {E}uropean network of protected sites} (2016).
\newline\urlprefix\url{http://www.eea.europa.eu/data-and-maps/data/natura-7}

\bibitem{Scholz}
Y.~Scholz, {Renewable energy based electricity supply at low costs -
  Development of the REMix model and application for {Europe}}, Ph.D. thesis,
  Universit\"at Stuttgart (2012).
\newblock \href {http://dx.doi.org/10.18419/opus-2015}
  {\path{doi:10.18419/opus-2015}}.

\bibitem{pfluger2011}
B.~Pfluger, F.~Sensfu{\ss}, G.~Schubert, J.~Leisentritt,
  \href{https://www.isi.fraunhofer.de/content/dam/isi/dokumente/ccx/2011/Final_Report_EU-Long-term-scenarios-2050.pdf}{Tangible
  ways towards climate protection in the {Europe}an {U}nion ({EU} long-term
  scenarios 2050)}, Tech. rep., Fraunhofer ISI (2011).
\newline\urlprefix\url{https://www.isi.fraunhofer.de/content/dam/isi/dokumente/ccx/2011/Final_Report_EU-Long-term-scenarios-2050.pdf}

\bibitem{dee2011}
D.~P. Dee, S.~M. Uppala, A.~J. Simmons, P.~Berrisford, P.~Poli, S.~Kobayashi,
  U.~Andrae, M.~A. Balmaseda, G.~Balsamo, P.~Bauer, P.~Bechtold, A.~C.~M.
  Beljaars, L.~van~de Berg, J.~Bidlot, N.~Bormann, C.~Delsol, R.~Dragani,
  M.~Fuentes, A.~J. Geer, L.~Haimberger, S.~B. Healy, H.~Hersbach, E.~V.
  H{\'{o}}lm, L.~Isaksen, P.~K{\aa}llberg, M.~K{\"{o}}hler, M.~Matricardi,
  A.~P. McNally, B.~M. Monge-Sanz, J.-J. Morcrette, B.-K. Park, C.~Peubey,
  P.~de~Rosnay, C.~Tavolato, J.-N. Th{\'{e}}paut, F.~Vitart, The {ERA-Interim}
  reanalysis: configuration and performance of the data assimilation system,
  Quarterly Journal of the Royal Meteorological Society 137~(656) (2011)
  553--597.
\newblock \href {http://dx.doi.org/10.1002/qj.828} {\path{doi:10.1002/qj.828}}.

\bibitem{Pfenninger2017}
S.~Pfenninger, {Dealing with multiple decades of hourly wind and {PV} time
  series in energy models: A comparison of methods to reduce time resolution
  and the planning implications of inter-annual variability}, Applied Energy
  197 (2017) 1--13.
\newblock \href {http://dx.doi.org/10.1016/j.apenergy.2017.03.051}
  {\path{doi:10.1016/j.apenergy.2017.03.051}}.

\bibitem{Kotzur2018}
L.~Kotzur, P.~Markewitz, M.~Robinius, D.~Stolten, {Impact of different time
  series aggregation methods on optimal energy system design}, Renewable Energy
  117 (2018) 474--487.
\newblock \href {http://dx.doi.org/10.1016/j.renene.2017.10.017}
  {\path{doi:10.1016/j.renene.2017.10.017}}.

\bibitem{Hartel2017}
P.~H{\"{a}}rtel, M.~Kristiansen, M.~Korp{\aa}s, {Assessing the impact of
  sampling and clustering techniques on offshore grid expansion planning},
  Energy Procedia 137 (2017) 152--161.
\newblock \href {http://dx.doi.org/10.1016/j.egypro.2017.10.342}
  {\path{doi:10.1016/j.egypro.2017.10.342}}.

\bibitem{jacob2014}
D.~Jacob, J.~Petersen, B.~Eggert, A.~Alias, O.~B. Christensen, L.~M. Bouwer,
  A.~Braun, A.~Colette, M.~D{\'e}qu{\'e}, G.~Georgievski, E.~Georgopoulou,
  A.~Gobiet, L.~Menut, G.~Nikulin, A.~Haensler, N.~Hempelmann, C.~Jones,
  K.~Keuler, S.~Kovats, N.~Kr{\"o}ner, S.~Kotlarski, A.~Kriegsmann, E.~Martin,
  E.~van Meijgaard, C.~Moseley, S.~Pfeifer, S.~Preuschmann, C.~Radermacher,
  K.~Radtke, D.~Rechid, M.~Rounsevell, P.~Samuelsson, S.~Somot, J.-F. Soussana,
  C.~Teichmann, R.~Valentini, R.~Vautard, B.~Weber, P.~Yiou, {EURO-CORDEX: new
  high-resolution climate change projections for {Europe}an impact research},
  Regional Environmental Change 14~(2) (2014) 563--578.
\newblock \href {http://dx.doi.org/10.1007/s10113-013-0499-2}
  {\path{doi:10.1007/s10113-013-0499-2}}.

\bibitem{DEANE2014152}
J.~Deane, G.~Drayton, B.~O. Gallach\'oir, The impact of sub-hourly modelling in
  power systems with significant levels of renewable generation, Applied Energy
  113 (2014) 152--158.
\newblock \href {http://dx.doi.org/10.1016/j.apenergy.2013.07.027}
  {\path{doi:10.1016/j.apenergy.2013.07.027}}.

\bibitem{6345631}
N.~Troy, D.~Flynn, M.~O'Malley, The importance of sub-hourly modeling with a
  high penetration of wind generation, in: 2012 IEEE Power and Energy Society
  General Meeting, 2012, pp. 1--6.
\newblock \href {http://dx.doi.org/10.1109/PESGM.2012.6345631}
  {\path{doi:10.1109/PESGM.2012.6345631}}.

\bibitem{ODwyer2015}
C.~O'Dwyer, D.~Flynn, Using energy storage to manage high net load variability
  at sub-hourly time-scales, IEEE Transactions on Power Systems 30~(4) (2015)
  2139--2148.
\newblock \href {http://dx.doi.org/10.1109/TPWRS.2014.2356232}
  {\path{doi:10.1109/TPWRS.2014.2356232}}.

\bibitem{burdenresponse}
T.~Brown, T.~Bischof-Niemz, K.~Blok, C.~Breyer, H.~Lund, B.~Mathiesen,
  {Response to `Burden of proof: A comprehensive review of the feasibility of
  100\% renewable-electricity systems'}, Renewable and Sustainable Energy
  Reviews 92 (2018) 834--847.
\newblock \href {http://dx.doi.org/10.1016/j.rser.2018.04.113}
  {\path{doi:10.1016/j.rser.2018.04.113}}.

\bibitem{etip}
E.~Vartiainen, G.~Masson, C.~Breyer,
  \href{http://www.etip-pv.eu/fileadmin/Documents/ETIP_PV_Publications_2017-2018/LCOE_Report_March_2017.pdf}{{The
  True Competitiveness of Solar {PV}: A {Europe}an Case Study}}, Tech. rep.,
  European Technology and Innovation Platform for Photovoltaics (2017).
\newline\urlprefix\url{http://www.etip-pv.eu/fileadmin/Documents/ETIP_PV_Publications_2017-2018/LCOE_Report_March_2017.pdf}

\bibitem{Nykvist2015}
B.~Nykvist, M.~Nilsson, {Rapidly falling costs of battery packs for electric
  vehicles}, Nature Climate Change 5~(4) (2015) 329--332.
\newblock \href {http://dx.doi.org/10.1038/nclimate2564}
  {\path{doi:10.1038/nclimate2564}}.

\bibitem{EUsurvey2017}
{European Commission}, {Special Eurobarometer 459 Report: Climate Change},
  Tech. rep., {European Commission} (2017).
\newblock \href {http://dx.doi.org/10.2834/92702} {\path{doi:10.2834/92702}}.

\bibitem{ellis_social_2016}
G.~Ellis, G.~Ferraro, The social acceptance of wind energy, {JRC} {Science} for
  {Policy} {Report} EUR 28182 EN, The Joint Research Centre (JRC) (2016).
\newblock \href {http://dx.doi.org/10.2789/696070} {\path{doi:10.2789/696070}}.

\bibitem{horsch_role_2017}
J.~H\"{o}rsch, T.~Brown, The role of spatial scale in joint optimisations of
  generation and transmission for {European} highly renewable scenarios, in:
  2017 14\textsuperscript{th} {International} {Conference} on the {European}
  {Energy} {Market} ({EEM}), 2017.
\newblock \href {http://dx.doi.org/10.1109/EEM.2017.7982024}
  {\path{doi:10.1109/EEM.2017.7982024}}.

\bibitem{EEACO22017}
{European Environment Agency},
  \href{https://www.eea.europa.eu/ds_resolveuid/d2a35821365d4ce09b277ab3a85bf305}{{National
  emissions reported to the UNFCCC and to the EU Greenhouse Gas Monitoring
  Mechanism}} (2017).
\newline\urlprefix\url{https://www.eea.europa.eu/ds_resolveuid/d2a35821365d4ce09b277ab3a85bf305}

\bibitem{schlachtberger_backup_2016}
D.~Schlachtberger, S.~Becker, S.~Schramm, M.~Greiner, Backup flexibility
  classes in emerging large-scale renewable electricity systems, Energy
  Conversion and Management 125 (2016) 336 -- 346.
\newblock \href {http://dx.doi.org/10.1016/j.enconman.2016.04.020}
  {\path{doi:10.1016/j.enconman.2016.04.020}}.

\bibitem{biggar2014economics}
D.~R. Biggar, M.~R. Hesamzadeh, The Economics of Electricity Markets, Wiley,
  2014.

\bibitem{bjelic2015}
I.~{Batas Bjeli{\'{c}}}, N.~Rajakovi{\'{c}}, {Simulation-based optimization of
  sustainable national energy systems}, Energy 91 (2015) 1087--1098.
\newblock \href {http://dx.doi.org/10.1016/j.energy.2015.09.006}
  {\path{doi:10.1016/j.energy.2015.09.006}}.

\bibitem{bjelic2016}
I.~{Batas Bjeli{\'{c}}}, N.~Rajakovi{\'{c}}, G.~Kraja{\v{c}}i{\'{c}},
  N.~Dui{\'{c}}, {Two methods for decreasing the flexibility gap in national
  energy systems}, Energy 115 (2016) 1701--1709.
\newblock \href {http://dx.doi.org/10.1016/j.energy.2016.07.151}
  {\path{doi:10.1016/j.energy.2016.07.151}}.

\bibitem{Lund2010}
H.~Lund (Ed.), The Choice and Modelling of 100\% Renewable Solutions, Academic
  Press, 2010.

\bibitem{Brown2018}
T.~Brown, D.~Schlachtberger, A.~Kies, M.~Greiner, {Synergies of sector coupling
  and transmission extension in a cost-optimised, highly renewable {Europe}an
  energy system }, Energy 160 (2018) 720--730.
\newblock \href {http://dx.doi.org/10.1016/j.energy.2018.06.222}
  {\path{doi:10.1016/j.energy.2018.06.222}}.

\bibitem{staffell_using_2016}
I.~Staffell, S.~Pfenninger, Using bias-corrected reanalysis to simulate current
  and future wind power output, Energy 114 (2016) 1224--1239.
\newblock \href {http://dx.doi.org/10.1016/j.energy.2016.08.068}
  {\path{doi:10.1016/j.energy.2016.08.068}}.

\bibitem{Hdidouan2017}
D.~Hdidouan, I.~Staffell, {The impact of climate change on the levelised cost
  of wind energy}, Renewable Energy 101 (2017) 575--592.
\newblock \href {http://dx.doi.org/10.1016/j.renene.2016.09.003}
  {\path{doi:10.1016/j.renene.2016.09.003}}.

\bibitem{Wohland2017}
J.~Wohland, M.~Reyers, J.~Weber, D.~Witthaut, More homogeneous wind conditions
  under strong climate change decrease the potential for inter-state balancing
  of electricity in {Europe}, Earth System Dynamics 8~(4) (2017) 1047--1060.
\newblock \href {http://dx.doi.org/10.5194/esd-8-1047-2017}
  {\path{doi:10.5194/esd-8-1047-2017}}.

\bibitem{Karnauskas2018}
K.~B. Karnauskas, J.~K. Lundquist, L.~Zhang, {Southward shift of the global
  wind energy resource under high carbon dioxide emissions}, Nature Geoscience
  11~(1) (2018) 38--43.
\newblock \href {http://dx.doi.org/10.1038/s41561-017-0029-9}
  {\path{doi:10.1038/s41561-017-0029-9}}.

\bibitem{Lund2003}
H.~Lund, Excess electricity diagrams and the integration of renewable energy,
  International Journal of Sustainable Energy 23~(4) (2003) 149--156.
\newblock \href {http://dx.doi.org/10.1080/01425910412331290797}
  {\path{doi:10.1080/01425910412331290797}}.

\bibitem{Collins2018}
S.~Collins, P.~Deane, B.~{\'{O}}. Gallach{\'{o}}ir, S.~Pfenninger, I.~Staffell,
  Impacts of inter-annual wind and solar variations on the {Europe}an power
  system, Joule, in press.
\newblock \href {http://dx.doi.org/10.1016/j.joule.2018.06.020}
  {\path{doi:10.1016/j.joule.2018.06.020}}.

\end{thebibliography}

\end{document}